\documentclass[lettersize,journal]{IEEEtran}
\usepackage{amsmath,amsfonts}
\usepackage{algorithm}
\usepackage{algpseudocode}
\usepackage{array}
\usepackage{textcomp}
\usepackage{url}
\usepackage{verbatim}
\usepackage{graphicx}
\usepackage{cite}
\usepackage{glossaries}
\usepackage{float}
\usepackage{subcaption}
\usepackage{dblfloatfix}

\algnewcommand\algorithmicforeach{\textbf{for each}}
\algdef{S}[FOR]{ForEach}[1]{\algorithmicforeach\ #1\ \algorithmicdo}

\usepackage{tikz}
\usepackage{pgfplots}
\usepackage{circuitikz}

\usetikzlibrary{patterns}
\usetikzlibrary{backgrounds}

\usetikzlibrary{decorations.pathreplacing,decorations.markings,shapes.geometric}
\tikzset{naming/.style={align=center,font=\small}}
\tikzset{antenna/.style={insert path={-- coordinate (ant#1) ++(0,0.25) -- +(135:0.25) + (0,0) -- +(45:0.25)}}}
\tikzset{station/.style={naming,draw,shape=dart,shape border rotate=90, minimum width=10mm, minimum height=10mm,outer sep=0pt,inner sep=3pt}}
\tikzset{mobile/.style={naming,draw,shape=rectangle,minimum width=8mm,minimum height=12mm, outer sep=0pt,inner sep=3pt}}

\newcommand{\BS}[1]{%
\begin{tikzpicture}
\node[station] (base) {#1};

\draw[line join=bevel] (base.100) -- (base.80) -- (base.110) -- (base.70) -- (base.north west) -- (base.north east);
\draw[line join=bevel] (base.100) -- (base.70) (base.110) -- (base.north east);

\draw[line cap=rect] ([yshift=0pt]base.north) [antenna=1];
\end{tikzpicture}
}

\newcommand{\UE}[1]{%
\begin{tikzpicture}[every node/.append style={rectangle,minimum width=0pt}]
\node [mobile,label={[inner ysep=+.3333em]\dots}] (box) {#1};

\draw ([xshift=3] box.north west) [antenna=1];
\draw ([xshift=-3]box.north east) [antenna=2];
\end{tikzpicture}
}

\usetikzlibrary{shapes}
\usetikzlibrary{calc}

\def \fwidth{0.8\linewidth}
\def \fheight{0.6\linewidth}

\def \cmwidth{0.8\linewidth}
\def \cmheight{0.8\linewidth}

\def \cbwidth{0.03\linewidth}

\def \cmfullwidth{0.75\linewidth}
\def \cmfullheight{0.75\linewidth}
\def \cbfullwidth{0.015\linewidth}

\colorlet{mycolor1}{black}
\colorlet{mycolor2}{cyan}
\colorlet{mycolor3}{orange}
\colorlet{mycolor4}{violet}
\colorlet{mycolor5}{red}

\newacronym{aoa}{AoA}{Angle of Arrival}
\newacronym{toa}{ToA}{Time of Arrival}
\newacronym[plural=ADCs,firstplural=Analog to Digital Converters]{adc}{ADC}{Analog to Digital Converter}
\newacronym{csi}{CSI}{Channel State Information}
\newacronym[plural=TRNs,firstplural=Training Fields]{trn}{TRN}{Training Field}
\newacronym{lpf}{LPF}{Low Pass Filter}
\newacronym{music}{MUSIC}{MUltiple SIgnal Classification}
\newacronym{2dmusic}{2D-MUSIC}{2D MUltiple SIgnal Classification}
\newacronym{mimo}{MIMO}{Multiple Input Multiple Output}
\newacronym{mumimo}{MU-MIMO}{Multi User Multiple Input Multiple Output}
\newacronym{simo}{SIMO}{Single Input Multiple Output}
\newacronym{cir}{CIR}{Channel Impulse Response}
\newacronym{cfr}{CFR}{Channel Frequency Response}
\newacronym{mrc}{MRC}{Maximum Ratio Combining}
\newacronym{ula}{ULA}{Uniform Linear Array}
\newacronym{ue}{UE}{User Equipment}
\newacronym{bs}{BS}{Base Station}
\newacronym{tdd}{TDD}{Time Division Duplexing}
\newacronym{ofdm}{OFDM}{Orthogonal Frequency Division Multiplexing}
\newacronym{af}{AF}{Array Factor}
\newacronym{snr}{SNR}{Signal to Noise Ratio}
\newacronym{los}{LoS}{Line of Sight}
\newacronym{nlos}{NLoS}{Non Line of Sight}
\newacronym{tx}{TX}{Transmitter}
\newacronym{rx}{RX}{Receiver}
\newacronym{rf}{RF}{Radio Frequency}
\newacronym{mu}{MU}{Multi User}
\newacronym{rms}{RMS}{Root Mean Square}
\newacronym{fft}{FFT}{Fast Fourier Transform}
\newacronym{wlan}{WLAN}{wireless local area network}
\newacronym{ofdma}{OFDMA}{Orthogonal Frequency Division Multiplexing Access}
\newacronym{jcs}{JCAS}{Joint Communication and Sensing}
\newacronym{pmcw}{PMCW}{phase modulated continuous wave}
\newacronym{mmwave}{mmWave}{millimeter wave}
\newacronym{rmse}{RMSE}{Root Mean Square Error}
\newacronym{iir}{IIR}{Infinite Impulse Response}
\newacronym{LFM}{LFM}{linear frequency modulated}
\newacronym{FMCW}{FMCW}{frequency modulated continuous wave}
\newacronym{PPM}{PPM}{pulse position modulated}
\newacronym{eurllc}{eURLLC}{extreme Ultra Reliable Low Latency Communications}
\newacronym{pec}{PEC}{Perfect Electric Conductor}
\newacronym{cst}{CST}{CST Microwave Studio}
\newacronym{dtft}{DTFT}{Discrete Time Fourier Transform}

\usepackage[normalem]{ulem}

\begin{document}

\title{Low-complexity hardware and algorithm for joint communication and sensing}

\author{Andrea Bedin,~\IEEEmembership{Graduate Student Member, IEEE}, Shaghayegh Shahcheraghi, Traian E. Abrudan,~\IEEEmembership{Member, IEEE}, Arash Asadi,~\IEEEmembership{Senior Member, IEEE}.
\thanks{
Andrea Bedin (andrea.bedin.2@studenti.unipd.it) and Traian E. Abrudan (traian.abrudan@nokia-bell-labs.com) are with Nokia Bell Labs, Espoo, Finland. Andrea Bedin is also  with the Department of Information Engineering, University of Padova, Italy. Arash Asadi (aasadi@wise.tu-darmstadt.de) is with the computer science department of TU Darmstadt. Shaghaeygh Shahcheraghi (shahcheraghis@gmail.com) was with the computer science department of TU Darmstadt from 2020 to 2022. This project has received funding from the European Union's Horizon 2020 research and innovation program under the Marie Skłodowska-Curie Grant agreement No. 861222.
}}

\maketitle

\begin{abstract}
\gls{jcs} is foreseen as one very distinctive feature of the emerging 6G systems providing, in addition to fast end reliable communication, the ability to obtain an accurate perception of the physical environment. 
In this paper, we propose a \gls{jcs} algorithm that exploits a novel beamforming architecture, which features a combination of wideband analog and narrowband digital beamforming. This allows accurate estimation of \gls*{toa}, exploiting the large bandwidth and \gls*{aoa}, exploiting the  high-rank digital beamforming. In our proposal, we separately estimate the \gls{toa} and \gls{aoa}. The association between \gls{toa} and \gls{aoa} is solved by acquiring multiple non-coherent frames and adding up the signal from each frame such that a specific component is combined coherently before the \gls{aoa} estimation. Consequently, this removes the need to use 2D and 3D joint estimation methods, thus significantly lowering complexity. The resolution performance of the method is compared with that of \gls{2dmusic} algorithm, using a fully-digital wideband beamforming architecture. The results show that the proposed method can achieve performance similar to a fully-digital high-bandwidth system, while requiring a fraction of the total aggregate sampling rate and having much lower complexity.\\ 
\end{abstract}


\glsresetall

\glsunset{mimo}
\glsunset{mumimo}
\glsunset{simo}
\glsunset{ofdm}
\glsunset{snr}
\glsunset{mmwave}
\glsunset{rf}

\section{Introduction} 
The primary differentiation of 6G compared to 5G is \gls{jcs}. There are two key enablers of \gls{jcs}. 
First, {\em the large bandwidth} available in the millimeter-wave spectrum enables not only higher data rates~\cite{hexax_6G_embb}, but also higher ranging resolution. 
Second, the shorter wavelengths allow for very compact {\em large aperture antenna arrays}, thus enabling high-resolution beamforming and angular estimation.
Sensing information can be used standalone, e.g., for user localization and navigation, and in imaging applications~\cite{hexax_6G, hexax_6G_2, DOPPLER2022101571}. In addition, it can be used to enhance communication through better beam selection and preventing disruptions caused by blockages which are persistent issues of millimeter-wave communication systems. 6G is expected to support  \gls{eurllc} applications \cite{hexax_6G_urllc, hexax_6G_urllc2}, where best-effort service does not suffice anymore. An example of such traffic is cooperative robots, which according to \cite{hexax_6G_urllc}, despite the low data rate on the order of kbps, may require failure rates as low as $10^{-9}$. Another notable use case is self-driving vehicles, which pose specific requirements including: low throughput \gls{eurllc} for vehicle coordination and safety features, e.g., emergency braking and collision avoidance; \gls{jcs} for obstacle detection and environment mapping; and Massive data rates for entertainment and cooperative sensing.

Since the key sensing parameters are the estimated \gls*{toa} and \gls*{aoa} corresponding to the targets of interest, we ideally need a high-rank and high-bandwidth full-\gls{mimo} systems. Although this is certainly possible theoretically, it poses major technical challenges in terms of power consumption and cost. Considering that a multi-GHz \gls*{adc} can have a power consumption of over 2 watts 
\cite{adc1, adc2}, large arrays using an individual \gls*{adc} for each antenna become power hungry and expensive devices which are deemed unfeasible for commercial cellular devices, especially on the \gls*{ue} side. For example, A UE with a $16$ element array connected to  $2$W \gls*{adc}s would consume $32$W only for analog to digital conversion, which is impractical.

Another relevant implementation challenge is the computational complexity of the sensing algorithms, which is a major practical limitation in the state-of-the-art sensing algorithms. Methods like \gls{2dmusic} are extremely complex due to the size of the covariance matrix and the search space. Finally, mono-static radar for sensing is practically infeasible, especially in mobile handsets, because they require very complex in-band full-duplex receivers~\cite{2016KoTaTu+}.

\subsection*{Contributions}

We note that bandwidth is not critical for \gls*{aoa} estimation, and the \gls{eurllc} traffic which can benefit from large rank is typically low-throughput. Similarly, the array size has very little impact on \gls*{toa} estimation, and massive data rates are also achievable with a low-rank system with analog beamforming. Therefore, we do not need to digitalize the signal from all antenna elements on the full bandwidth. We propose a novel hardware architecture that adds, on top of the usual analog beamforming, an individual narrowband \gls*{rf} chain for each antenna. The signal from the individual \gls{rf} chains is then multiplexed into a single \gls{adc} and digitalized using a fraction of the sampling rate of the \gls{adc} for each antenna. 

More precisely, in this article, we make the following technical contributions:

\begin{itemize}

\item[C1)] In Section~\ref{sec:arch}, we propose {\em a novel hardware architecture} for low-complexity high-resolution sensing. The proposed architecture significantly reduces the power consumption and cost of the mmWave \gls*{jcs} system by combining an equivalent network of low-bandwidth \glspl*{adc} (realized by multiplexing multiple signals into a single \gls{adc}) for digital beamforming with a high-bandwidth \gls*{adc} for analog beamforming. In addition, we propose a modulation scheme and beam design mechanism that is suitable for both communication and sensing, and provide a mathematical description of the output of the channel estimate that will be used later for \gls{aoa} and \gls{toa} estimation. 

\item[C2)] 
In Section~\ref{sec:sens}, we propose a novel \gls{jcs} algorithm exploiting the proposed architecture. First, using the wideband analog beamformer of the architecture, \gls*{toa} is accurately estimated (e.g. using the \gls*{music} algorithm). Then, \gls*{mrc} is appied in the digital beamforming domain on multiple non-coherent frames to amplify the path component associated to each \gls*{toa}. Finally, the corresponding \gls*{aoa} is estimated exploiting the combined digital beamforming signal (e.g. using Matrix Pencil). This way, \gls*{aoa} and \gls*{toa} of the multipath components are estimated with high resolution and low complexity. The theoretical advantages of the proposed architecture and method are explored in Section~\ref{sec:theoretical_analysis}.

\item[C3)] In Section~\ref{sec:results}, we evaluate the performance of the proposed method, both in terms of parameter estimation error and close target resolution capabilities, and compare its performance with the performance of \gls{2dmusic}. We show that, despite the dramatically lower hardware and software complexity and the reduced power consumption, the proposed system has comparable performance to state-of-the-art solutions. 

\end{itemize} 

\subsection*{Advantages of the Proposed Architecture and Method} 

The proposed architecture possesses several advantages, both for communication and sensing, as explained below.

\begin{itemize}

\item[A1)] {\em Versatility:}
The proposed reconfigurable architecture comprising both a high-bandwidth analog beamformer and a low-bandwidth digital beamformer can support the vast range of requirements of modern standards much better than a classic analog or hybrid beamforming solution, while being considerably less expensive than a fully digital \gls*{mimo} system.

\item[A2)] {\em One-shot beam scanning:} 
Our architecture overcomes the limitations of codebook-based beamforming approaches used in communication \cite{MRC1, MRC2} by designing the beam in real-time using \gls*{mrc}. Such a beam design procedure, in fact, requires individual knowledge of the channel for each antenna. Acquiring such information in a classical analog beamforming system imposes a significant overhead, as we are required to sweep through the codebook every time an updated channel state information is needed. This limits the frequency of the beam update, which in a very dynamic environment could lead to momentary disruptions in communication. As in the case of sensing though, it might not be necessary to sample all the antennas in the full bandwidth to obtain the channel state information required to design the beam. In our architecture, the beam can be thus optimized very often based on the narrowband digital beamforming sub-system channel estimates, while removing the necessity of time-consuming beam sweep procedures. It should be also noted that this beam update can be done {\em without interrupting the communication}, meaning that we remove any overhead related to beam training, at the expense of a slightly higher power consumption at the front end.

\item[A3)] {\em Robust Low-SNR Sensing:} 
The use of analog beamformer for the \gls{toa} estimate provides enhanced \gls{snr}, thus preventing the \gls{snr} collapse \cite{MUSIC_collapse} of the \gls{music} algorithm, i.e. the situation where a noise eigenvector is selected over a signal one due to random fluctuations of the eigenvalues. 

\item[A4)]  {\em Low Computational Complexity:} 
Finally, the method relies on subsequent 1D parameter estimation, rather than joint 2d estimations. This greatly improves its computational complexity.

\end{itemize}

\section{related work} \label{sec:related}

Most of the recent literature integrated radar sensing with communication where a single waveform is used to extract both target parameters and communication symbols. The choice of waveform has a significant effect on the performance of \gls{jcs}. The existing \gls{jcs} waveforms may be classified into radar-based waveforms, communication-based waveforms and optimised \gls {jcs} waveforms\cite{ofdm1,ofdm2,ofdm3,ofdm4,uchidaSS,rSTANDARDdemonstrating,NOISEOFDMuwb,OFDM2006combining, garmatyukOFDMwideband, dokhanchi2018ofdm, sturm2013spectrally}. In the following, we provide an overview of these works.

{\bf  \gls*{jcs} using radar waveforms.} Some prior works embed communication information in typical radar waveforms such as, \gls*{FMCW} \cite{fmcw}, \gls*{pmcw} \cite{gianninipmcw, guermandipmcw, dokhanchipmcwjoint},  \gls*{LFM} wave \cite{zhangLFMwaveform}, and \gls*{PPM} wave \cite{pulsepositionMode}. The general drawback of these radar-based \gls*{jcs} approach is that they cannot provide high data-rate transmission. 
In another approach, \cite{pmcwdirectss} proposes a \gls*{pmcw} with a direct sequence spread-spectrum technique to achieve high throughput for \gls*{jcs}, where there is a trade-off between the length of the code sequence and data rate. Also, since it uses large code sequences, it suffers from high complexity and low energy efficiency.

{\bf  \gls*{jcs} using communication waveforms.} \gls{ofdm} is one of the most common communication waveforms, and therefore is also widely used for \gls*{jcs}  \cite{ofdm1,ofdm2,ofdm3,ofdm4,OFDM2006combining, garmatyukOFDMwideband, dokhanchi2018ofdm}. Some works use standard compliant communication waveforms for \gls*{jcs} \cite{ofdm3,robertieeead, kumari2015investigating, grossi2018opportunistic, kumari2017ieee, muns2019beam, mishra2017sub,storrer2021indoor}. For example, \cite{ofdm3} uses IEEE 802.11p car-to-car communication standard for automotive radar, and \cite{storrer2021indoor} uses 802.11ax Wi-Fi-based for indoor human tracking, but due to the small bandwidth at sub-6GHz frequencies, the communication and sensing performance are limited. To overcome this issue, in \cite{robertieeead}, IEEE 802.11ad \gls*{mmwave} \gls*{wlan} standard is exploited to realize a joint waveform for long-range radar applications. However, its performance is limited by the channel estimation methods implemented in the communication protocol. Moreover, a full-duplex radar operation is assumed, which makes the hardware implementation infeasible due to its complex self-interference cancellation requirement. 
Spread spectrum \cite{uchidaSS} and noise-\gls*{ofdm} \cite{NOISEOFDMuwb} are other communication waveforms that have been used for \gls*{jcs}.

{\bf  \gls*{jcs} using optimised waveforms.} Another option is to design a waveform that is suitable for \gls*{jcs} by optimizing the metrics of radar and communication performance \cite{liu2017adaptive, kumari2019adaptive}. In general, finding the optimal waveform can be computationally expensive.

Contrary to these works, we propose a novel architecture that enables \gls*{jcs} with high resolution, low complexity and low power consumption. In this paper, the received signals that are reflected from different vehicles are processed in a bi-static OFDM-based scenario, and the sensing parameters are estimated using the proposed method.

\section{System model} \label{sec:sysmod}

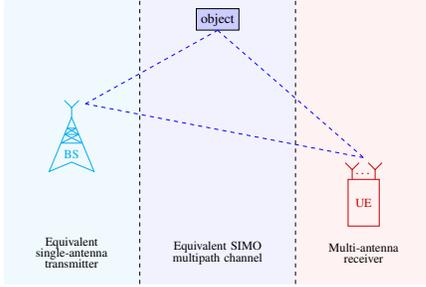
\begin{figure}[t]
\centering
\resizebox{0.65\linewidth}{!}{\begin{tikzpicture}

\node[cyan](bs) at (0.25, 0){\BS{BS}};

\node[draw, black, fill=blue!20!white](obj) at (4,3) {object};

\node[red!80!black](ue) at (7.75, -1.5) {\UE{UE}} ;

\draw[dashed, blue] ([xshift=10,yshift=-6]bs.north) -- (obj.south);

\draw[dashed, blue] ([xshift=10,yshift=-6]bs.north) -- (ue.north) ;

\draw[dashed, blue] (obj.south) -- (ue.north) ;

\draw[dashed, black] (2,-3.8) -- (2,3.5);

\draw[dashed, black] (6,-3.8) -- (6,3.5);

\node[text width = 80, align = center, font=\small\linespread{0.8}\selectfont] at (0.25, -3) {Equivalent\\ single-antenna \\transmitter};

\node[text width = 80, align = center, font=\small\linespread{0.8}\selectfont] at (4, -3) {Equivalent SIMO multipath channel};

\node[text width = 80, align = center, font=\small\linespread{0.8}\selectfont] at (7.75, -3) {Multi-antenna\\ receiver};

\begin{scope}[on background layer]
\fill[cyan!5!white] (-1.5,-3.8) rectangle (2,3.5); 

\fill[blue!5!white]  (2,3.5) rectangle (6,-3.8); 

\fill[red!5!white]  (6,-3.8) rectangle (9.5, 3.5); 
\end{scope}

\end{tikzpicture}
 }
\caption{System overview.}
\label{fig:sys_overview}
\end{figure}

In this paper, we consider a system as depicted in Fig. ~\ref{fig:sys_overview}, where a \gls*{bs} transmits data to a \gls*{ue} in a multipath environment. The \gls*{ue} receives the signal transmitted by the \gls*{bs}, estimates the \gls{csi} and estimate the \gls{aoa} and \gls{toa} of the received multipath components. Although the \gls*{bs} could be using \gls*{mumimo}, for simplicity, in this analysis, we only consider a single beamformed stream directed to the \gls*{ue} of interest. Consequently, we can use a \gls{simo} channel model, and consider any interference from other \gls{mumimo} streams as part of the \gls{snr}. Moreover, we assume that either the \gls{ue} or the environment are not static (e.g., vehicular scenarios), and therefore the phases of the multipath components are not constant over the acquisition of multiple frames due to the Doppler shift. For example, at a frame rate of $0.1$ms, the position of a car moving at $15$m/s ($54$ kmh) is displaced by $1.5$mm. Although negligible in terms of range, at $60$GHz carrier frequency, this displacement introduces a phase shift larger than $90^{\circ}$.

 We assume that the received signal for frame $i$ can be decomposed in a sum of $K$ plane waves, each of which has an associated complex amplitude $\alpha_{(k,i)}$, delay $\tau_k$ and azimuth incidence angle $\theta_k$. We consider a \gls*{ula} with $N$ antennas spaced $\frac{\lambda}{2}$, where $\lambda$ is the wavelength of the received signal, and therefore do not consider the elevation angle. Nevertheless, the work can be adapted for more complex geometries. Finally, we assume that the channel is measured at subsequent time instants, and between those instants the geometry  stays the same (i.e. $\tau_k$ and $\theta_k$ are constant) but the channel coefficients $\alpha_{(k,i)}$ can change. 
With the above, the \gls*{cir} between the transmitter and antenna $n$ can be written as
\begin{equation}
h_n(i,t) = \sum_{k=0}^{K-1} \alpha_{(k,i)} e^{j  n \pi \cos(\theta_k) } \delta(t - \tau_k). 
\end{equation}
This definition leads to a baseband \gls*{cfr} of 
\begin{equation}
H_n(i,f) = \sum_{k=0}^{K-1} \alpha_{(k,i)} e^{j  n \pi \cos(\theta_k) } e^{-j 2 \pi f \tau_k} . \label{eq:cfr}
\end{equation}
with $f \in (-\frac{B_A}{2}, \frac{B_A}{2})$ and $B_A$ is the total bandwidth of the system.

\section{Hybrid \gls*{jcs} achitecture} \label{sec:comms}

In this section, we discuss the hardware architecture of the proposed system and briefly elaborate on its potential of enhancing communication. In addition, we propose a modulation scheme and beam design mechanism that is suitable for both communication and sensing, and provide a mathematical description of the output of the channel estimate that will be used later for \gls{aoa} and \gls{toa} estimation.

\subsection{Receiver hardware architecture} \label{sec:arch}

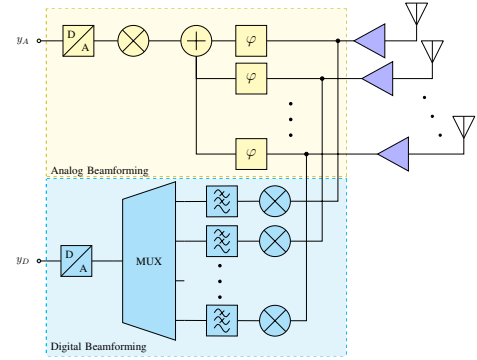
\begin{figure}[t] 
\centering
\resizebox{0.7\linewidth}{!}{\begin{tikzpicture}

\def \apos {-7}
\def \ashift {0.5}
\def \posend {-12}
\def \shiftapos {\apos + 0.5}
\def \mixpos {-9}
\def \shiftmixpos {\mixpos - 0.5}
\def \dmixpos {-4.5}
\def \ampendpos {-3}
\def \antpos {0}
\def \donepos {-2.5}
\def \dtwopos {-3}
\def \dthreepos {-3.5}
\def \lineone {0}
\def \linetwo {-1.2}
\def \linethree {-3.6}

\def \aone {\lineone + \ashift}
\def \atwo {\linetwo + \ashift}

\def \athree  {\linethree + \ashift}

\def \dlineone {-5.1}
\def \dlinetwo {-6.37}
\def \dlinethree {-8.9}

\def \posouttext {\posend - 0.5}

\tikzset{m1/.style={muxdemux, muxdemux def={Lh=7, Rh=9, NL=1, NB=0, NR=4}}}

\draw

node[adder, fill=yellow!30!white] at (\apos,0) (add) {}

node[bareantenna] (ant1) at (\antpos , \aone) {}
(ant1.center)  to node[short]{} (\antpos,  \lineone)
to [amp, fill=blue!30!white] node[short]{}(\ampendpos,\lineone) 
to node[short]{}(-4,\lineone) 
to [phaseshifter, fill=yellow!30!white] (add.east) 

node[bareantenna] (ant2)  at (\antpos + 0.5, \atwo) {}
(ant2.center)  to node[short]{} (\antpos+ 0.5,  \linetwo)
to [amp, fill=blue!30!white] node[short]{}(\ampendpos,\linetwo) 
to node[short]{}(-4,\linetwo) 
to [phaseshifter, fill=yellow!30!white] node[short]{}(\shiftapos,\linetwo)
to node[short]{}( \apos,\linetwo)
 to (add.south)
 
 node[bareantenna] (ant3)  at (\antpos + 1.5,  \athree) {}
(ant3.center)  to node[short]{} (\antpos+ 1.5,  \linethree)
to [amp, fill=blue!30!white] node[short]{}(\ampendpos, \linethree) 
to  node[short]{}(-4, \linethree) 
to [phaseshifter, fill=yellow!30!white] node[short]{}(\shiftapos, \linethree)
to node[short]{}( \apos, \linethree)
 to (add.south)
 
 node[mixer, fill=yellow!30!white] at (\mixpos,0) (mixwb) {}
 (add.west) to (mixwb.east)
 (mixwb.west) to[adc, -o, fill=yellow!30!white] node[short](analogend){}(\posend,0)
 
 
 
 node[m1, fill=cyan!30!white] (mux1) at (-8.5, -7) {MUX}
 
  node[mixer, fill=cyan!30!white] at (\dmixpos,\dlineone) (mixer1) {}
(\donepos,\lineone) to[short, *-] node[short]{} (\donepos,\dlineone)
to  (mixer1.east)

(mixer1.west) to [lowpass, fill=cyan!30!white] (mux1.rpin 1)

  node[mixer, fill=cyan!30!white] at (\dmixpos,\dlinetwo) (mixer2) {}
(\dtwopos,\linetwo) to[short, *-] node[short]{} (\dtwopos,\dlinetwo)
to  (mixer2.east)

(mixer2.west) to [lowpass, fill=cyan!30!white]  (mux1.rpin 2)

  node[mixer, fill=cyan!30!white] at (\dmixpos,\dlinethree) (mixer3) {}
(\dthreepos, \linethree) to[short, *-] node[short]{} (\dthreepos,\dlinethree)
to  (mixer3.east)

(mixer3.west) to [lowpass, fill=cyan!30!white]  (mux1.rpin 4)

 (mux1.lpin 1) to[adc, -o, fill=cyan!30!white] node[short](analogend){}(\posend,-7)


 ;

 \filldraw[color=black, fill=black, thick] (0.25,-1.8) circle (1pt) ;
\filldraw[color=black, fill=black, thick] (0.5,-2.4) circle (1pt) ;
 \filldraw[color=black, fill=black, thick] (0.75,-3) circle (1pt) ;

\filldraw[color=black, fill=black, very thick] (-4,-2) circle (1pt) ;
\filldraw[color=black, fill=black, very thick] (-4,-2.5) circle (1pt) ;
 \filldraw[color=black, fill=black, very thick] (-4,-3) circle (1pt) ;
 
 \filldraw[color=black, fill=black, very thick] (-6.25,-7.15) circle (1pt) ;
\filldraw[color=black, fill=black, very thick] (-6.25,-7.65) circle (1pt) ;
 \filldraw[color=black, fill=black, very thick] (-6.25,-8.15) circle (1pt) ;
 
\node at (\posouttext, \lineone) {$y_A$};
\node at (\posouttext, -7) {$y_D$};

\begin{scope}[on background layer]
	\draw[cyan!80!white, dashed, line width = 2] (-11.75, -4.4) -- (-2.25, -4.4) -- (-2.25, -10) -- (-11.75, -10) -- 	cycle;
	
	\fill[cyan!10!white] (-11.75, -4.4) -- (-2.25, -4.4) -- (-2.25, -10) -- (-11.75, -10);

	\draw[yellow!80!black, dashed, line width = 2] (-11.75, -4.3) -- (-2.25, -4.3) -- (-2.25,1) -- (-11.75, 1) -- cycle;
	
	\fill[yellow!10!white] (-11.75, -4.3) -- (-2.25, -4.3) -- (-2.25, 1) -- (-11.75, 1);
\end{scope}

  \node[anchor=west] at (-11.75,-9.75) 
    {Digital Beamforming};

  \node[anchor=west] at (-11.75,-4.15) 
    {Analog Beamforming};
 
\end{tikzpicture}}
\caption{Hardware architecture. The wideband analog beamforming and the narrowband fully-digital beamforming blocks are highlighted in yellow and blue, respectively.
}
\label{fig:architecture}
\end{figure}

To obtain good estimates for both the \gls{toa} and \gls{aoa}, we require a system that is wideband and high rank, while maintaining practical costs and power consumption. While fully digital \gls*{mimo} satisfies the former constraints, it fails on the last: its implementation is inherently costly and power-hungry due to the need for multiple high-speed \glspl{adc}. 
Our proposed architecture, depicted in Fig.~\ref{fig:architecture}, can achieve these goals. 
 It comprises of $N$ antennas connected to a classic analog beamforming chain (marked in yellow in the figure) that operates on the full bandwidth $B_A$ of the system. In particular, the signal from each antenna is amplified, phase shifted and added together in the analog domain. Subsequently, the combined signal is downconverted and digitalized by a single high-speed \gls{adc}. Calling the sampling period of the \gls{adc} $T \leq \frac{1}{B_A}$, the transmitted and received signal $x(t)$ and $y_A(t)$ respectively, and $\beta_n$ a complex beamforming coefficient with the amplitude determined by the LNA gain and the phase determined by the phase shifter, the output of the \gls{adc} would therefore be of the form
\begin{equation}
y_A(sT) = \sum_{n=0}^{N-1} \beta_n (h_n \ast x)(sT), \: s \in \mathbb{Z},
\end{equation}
where "*" denotes the convolution operator, and its \gls{dtft}, assuming that the signal is band-limited with bandwidth $B_A$, is
\begin{equation}
Y_A(f) = \sum_{n=0}^{N-1} \beta_n H_n X(f)
\end{equation}
On top of the calssical system, for each antenna, we implement an additional dedicated \gls*{rf} chain (marked in light blue) that has a narrower bandwidth. In particular, the signal from each antenna can be extracted after the low noise fronted amplifiers, individually downconverted to baseband and filtered with a \gls*{lpf}. We now re-purpose one \gls{adc} and multiplex all the low bandwidth individual antenna signals into that it to obtain a low bandwidth digital beamforming chain.
Let us define the set of antennas $\mathcal{M} = \{m_0,...,m_{M-1}\}, M \leq N$ for which we are enabling the individual \gls*{rf} chain. 
At the $s$-th sample of the \gls{adc}, we connect the multiplexer to the \gls{rf} chain of antenna $m_{(s  \mod  M)}$. With this assumption, and defining the filter impulse response as $g(t)$ and its transfer function as $G(f)$, the signal $y_D$ at the \gls{adc} output can be expressed as
\begin{equation}
y_D(kT) = (h_{m_{(s  \bmod  M)}} \ast g \ast  x)(sT), \: s \in \mathbb{Z}.
\end{equation}
From this signals, we can extract the individual signals 
\begin{align}
y_{j}(MsT) &= y_D((Ms + j)T)\\
& = (h_{m_{((Ms + j)  \bmod  M)}} \ast g \ast  x)((Ms + j)T)
\\
&=(h_{m_{j}} \ast g \ast  x)(MsT + jT),\\ &\: s \in \mathbb{Z}, \: j \in \{0,...,M-1\}\nonumber
\end{align}
Notably, the signal $y_{j}(MsT)$ includes all the contribution of antenna $m_{j}$ only, which is sampled regularly with a sampling interval $MT$. The signal is also delayed by a time $jT$, however, we ignore this shift as it can be compensated for in the digital domain. 
To satisfy the Nyquist–Shannon sampling theorem each of the signals must have a bandwidth of $B_D \leq \frac{B_A}{M}$. If we further assume the frequency response of the filter is 
\begin{equation}
 G(f) = \begin{cases}
     1 & \text{if} \: f \in \left( -\frac{B_D}{2}, \frac{B_D}{2} \right)\\
     0 & \text{elsewhere}
 \end{cases},
\end{equation}
 we can write the \gls{dtft} of the received signal as
\begin{equation}
Y_{j}(f) = H_{m_{j + 1}}(f)  X(f)
\end{equation}
for $f \in \left( -\frac{B_D}{2}, \frac{B_D}{2} \right)$.

The key fact there is that with this architecture, the total sampling rate required is twice as much as a purely analog system, but still $\frac{2}{N}$ times lower than a fully digital beamforming system, while maintaining both a full rank (though in a narrower band) and the full bandwidth. As an example, if we consider a mmWave 5G system with $400$MHz of bandwidth and 16 antennas, the total sampling rate required for classic analog beamforming would be $400$MS/s. By contrast, for the proposed architecture it would be $800$MS/s and for the fully digital architecture $6.4$GS/s. Scaling the power consumption of the \gls{adc} described in \cite{adc1} by the required bandwidth would correspond to $150$mW for the analog beamforming, $300$mW for the proposed architecture and $2.5$W for the fully digital \gls{mimo}. Therefore, by replacing the fully digital \gls{mimo} system with the proposed architecture we could achieve a power consumption reduction of $87.5\%$, which corresponds to roughly $2$W. This reduction is quite significant in a battery-operated device, as a $2$W power consumption alone (i.e. without considering the CPU, screen, etc..) could drain a typical smartphone battery (e.g. with a $15$Wh capacity) in $7.5$ hours.

\subsection{\gls{jcs} Key Aspects}

As mentioned in the introduction, despite this paper being focused on sensing, we want to briefly elaborate on the potential of this architecture in communications. In particular, we want to highlight its ability to support both \gls{eurllc} traffic and high data rates at the same time. Some possible strategies to achieve this goal are listed in the following:
\begin{itemize}
\item Using the digital beamforming part to obtain \gls{csi} and improve the analog beam design. This application is particularly relevant in situations where the channel varies quickly. In fact, classical beam sweep imposes major overhead in such situations, whereas when measuring the \gls{csi} with narrowband digital beamforming there is no overhead.
\item Using the analog beamforming part for high-throughput non-critical communications, such as video streaming, and the digital beamforming for low throughput critical time-sensitive transmissions, such as packets used for closed-loop control of machinery. The idea behind this implementation is to avoid packet loss due to beam failure: despite being a rare event, in an ultra-reliable setting with, e.g., a $99.9999\%$ reliability requirement, beam failure can become a likely enough event to prevent the system to meet its specifications.
\end{itemize}
We assume that \gls{csi} is inherently acquired by the communication system, e.g., for channel equalization and beamforming, so that this information is already available, and can be exploited for sensing. More specifically, we expect the sensing to have zero overhead on the radio channel, and do not cause any disruption to the communication.

\subsection{Modulation and analog beamforming}

For this analysis, we assume that the system uses \gls*{ofdm} with $S$ subcarriers\footnote{Since the FFT size is usually even (a power of two), we assume that the subcarrier with index -S/2 is a null subcarrier, such that the zero index subcarrier is located in the middle of the band.} and a subcarrier spacing of $\Delta_f$, thus the bandwidth is $B_A = S \Delta_f$. Such configuration is often found, e.g., in 4G and 5G commercial systems, and is foreseen to be maintained in 6G.

We assume that the transmitter includes some reference symbols in the transmitted signal. For each frame $i$, we obtain a noisy channel estimate by demodulating the signal and extracting such reference symbols. For the analog beamformer, as we are observing the signal $y_A(sT)$, we have the estimate of the channel after the phase shift and addition ,which can be expressed as:
\begin{eqnarray}
\hat{H}_A(i,s \Delta_f) = \sum_{n=0}^{N-1} \beta_{n,i} H_n(i,s \Delta_f) + w, \\ s \in \left\{-\left\lceil \frac{S}{2} \right\rceil+1, ..., \left\lfloor \frac{S}{2} \right\rfloor-1 \right\},
\end{eqnarray}
where $\beta_{n,i}$ is the complex beamforming coefficient for antenna $n$ and $w \sim \mathcal{N}(0, \sigma_n)$ is a white Gaussian noise. Similarly, for the digital beamforming we have the channel estimate for each antenna, obtained by demodulating the signals $y_D(MsT)$, which is:
\begin{eqnarray}
\hat{H}_m(i,s \Delta_f) = H_m(i, s \Delta_f) + w, \\ s \in \left\{-\left\lceil \frac{S}{2M} \right\rceil +1, ..., \left\lfloor \frac{S}{2M} \right\rfloor -1 \right\}, m \in \mathcal{M}.
\end{eqnarray}
If we assume that the beam is generated based on the available \gls{csi}, and is designed according to \gls{mrc} with an \gls{iir} filter to mitigate the noise of the estimate, at every frame the beamforming coefficients for the antennas in $\mathcal{M}$ are updated to:
\begin{equation}
\beta_{m,i+1} = \mu \beta_{m,i} + (1- \mu) \hat{H}^*_m(i,0),
\end{equation}
where $\mu$ is the \gls{iir} filter coefficient. Assuming that the channel at every antenna is updated regularly, after the initial beam alignment, this method allows to maintain a \gls{mrc} beam without requiring any beam sweep or other procedures.

\section{Theoretical advantage} \label{sec:theoretical_analysis}

In order to evaluate the potential \gls*{aoa} and  \gls*{toa} resolution capabilities of the aforementioned architecture, we use the analysis proposed in \cite{music_resolution_ToA}. We assume to acquire $F$=10 channel measurements on $N=16$
antennas with a subcarrier spacing of $\Delta_f$ = $400$~kHz, and we have a \gls*{snr} of $\Gamma_0$ = 10~dB when transmitting on $S$=1000 subcarriers. The \gls*{snr} $\Gamma$ will be scaled accordingly for systems that use less bandwidth to maintain the same transmitted power. With these definitions, we can rewrite the equations in \cite{music_resolution_ToA} to compute the \gls*{toa} resolution $\delta_T$ as 
\begin{equation}
\delta_T = \frac{1}{\pi  S \Delta_f} \sqrt[4]{ \frac{360 (S-2)}{\Gamma S F N} }. \label{eq:toa_res}
\end{equation}
To evaluate the \gls*{aoa} resolution, we consider two paths with distinct angles $\vartheta_1$ and $\vartheta_2$. In order to reuse the expression in Eq.~\eqref{eq:toa_res} for angular resolution, we define the spatial frequencies $\omega_1 = \frac{\cos(\vartheta_1)}{2}$ and $\omega_2 = \frac{\cos(\vartheta_2)}{2}$. In the spatial domain, the antenna aperture plays the role of the bandwidth in the time domain.
We thus have that the received phases associated with each component, as a function of the antenna number, are $2n \pi \omega_1 $ and $2n \pi \omega_2$. Since our antenna spacing is $\frac{\lambda}{2}$, we can write
\begin{equation}
\delta_\omega = \frac{2}{\pi  N } \sqrt[4]{ \frac{360 (N -2)}{\Gamma N F S} }.
\end{equation}

In order to quantify the gain that could be obtained with the proposed architecture, we compare the following systems:

{\bf Proposed system} uses the proposed architecture with $N$=16 antennas, $\vert \mathcal{M} \vert = N$, $B_A$ = $400$~MHz and $B_D$ = $25$~MHz. For this system, the transmission takes place across $S = 1000$ subcarriers so the \gls*{snr} is $\Gamma = \Gamma_0$.

{\bf Equivalent classical system} A conventional digital \gls*{mimo} architecture with $N' \in [4,16]$ antennas and a bandwidth of $\frac{2 B_A}{N'}$~GHz.  This system has a total sampling rate equivalent to the proposed system, but it only transmits on a fraction of the bandwidth, so its SNR is $\Gamma = \frac{N'}{2} \Gamma_0$.

{\bf  Full \gls{mimo} system} A conventional digital \gls*{mimo} architecture with $N =16$ antennas and $400$~MHz of bandwidth. The same \gls*{snr} consideration mentioned for the proposed system holds. It should be noted that the Full \gls{mimo} system is clearly much more expensive than the others as it requires a total of $16$ \gls*{adc}s with $400$~MHz bandwidth each. As such, it will be used as a baseline for comparison.

For the proposed system we use the analog beamforming part to compute the \gls*{toa} and the digital beamforming to compute the \gls*{aoa}.
 
\begin{figure*}[t]
     \centering
     \begin{subfigure}[t]{0.48 \linewidth}
%
%
\begin{tikzpicture}

\begin{axis}[%
width=\fwidth,
height=\fheight,
scale only axis,
xmin=4,
xmax=16,
xlabel style={font=\color{white!15!black}},
xlabel={$N'$},
xlabel near ticks,
ymin=0.2,
ymax=4.5,
ylabel style={font=\color{white!15!black}},
ylabel={ToA resolution [ns]},
ylabel near ticks,
axis background/.style={fill=white},
legend style={legend cell align=left, align=left, draw=white!15!black, at={(0.02,0.98)}, legend pos=north west, font=\footnotesize}
]

\addplot [color=red, mycolor1]
  table[row sep=crcr]{%
4	1.09559084993302\\
16	1.09559084993302\\
};
\addlegendentry{Proposed system}

\addplot [color=mycolor2, mark=x, mark options={solid, mycolor2}]
  table[row sep=crcr]{%
4	1.09504154374386\\
5	1.36845822546449\\
6	1.64173711449347\\
7	1.91487805468903\\
8	2.18788088954216\\
9	2.46074546217541\\
10	2.7334716153416\\
11	3.00605919142249\\
12	3.27850803242754\\
13	3.55081797999261\\
14	3.8229888753786\\
15	4.09502055947022\\
16	4.36691287277463\\
};
\addlegendentry{Equivalent classical system}

\addplot [color=mycolor3, dashed]
  table[row sep=crcr]{%
4	0.547795424966509\\
16	0.547795424966509\\
};
\addlegendentry{Full MIMO system}

\end{axis}

\end{tikzpicture}%
         \caption{Time resolution.}
         \label{fig:theor_ToA}
     \end{subfigure}
     \begin{subfigure}[t]{0.48 \linewidth}       
%
%
\begin{tikzpicture}

\begin{axis}[%
width=\fwidth,
height=\fheight,
scale only axis,
xmin=4,
xmax=16,
xlabel style={font=\color{white!15!black}},
xlabel={$N'$},
xlabel near ticks,
ymin=8,
ymax=28,
ylabel style={font=\color{white!15!black}},
ylabel={Spatial freq. resolution [mrad/ant]},
ylabel near ticks,
axis background/.style={fill=white},
legend style={legend cell align=left, align=left, draw=white!15!black, font=\footnotesize}
]

\addplot [color=mycolor1, dashdotted]
  table[row sep=crcr]{%
4	18.8524489924063\\
16	18.8524489924063\\
};
\addlegendentry{Proposed system}

\addplot [color=mycolor2, mark=x, mark options={solid, mycolor2}]
  table[row sep=crcr]{%
4	27.5664447710896\\
5	25.8060220492888\\
6	24.1863210618141\\
7	22.7817574971993\\
8	21.5719088193371\\
9	20.5239468592188\\
10	19.6083364577615\\
11	18.801147953874\\
12	18.0834928610044\\
13	17.4404991614876\\
14	16.8603869797904\\
15	16.3337403439838\\
16	15.8529567764679\\
};
\addlegendentry{Equivalent classical system}

\addplot [color=mycolor3, dashed]
  table[row sep=crcr]{%
4	9.42622449620314\\
16	9.42622449620314\\
};
\addlegendentry{full MIMO system}

\end{axis}

\end{tikzpicture}%
         \caption{Angular resolution.}
         \label{fig:theor_AoA}
     \end{subfigure}
        \caption{Theoretical resolution achievable by the proposed architecture and some classical fully digital beamfroming architectures using \gls{music}}
        \label{fig:theor_res}
\end{figure*}
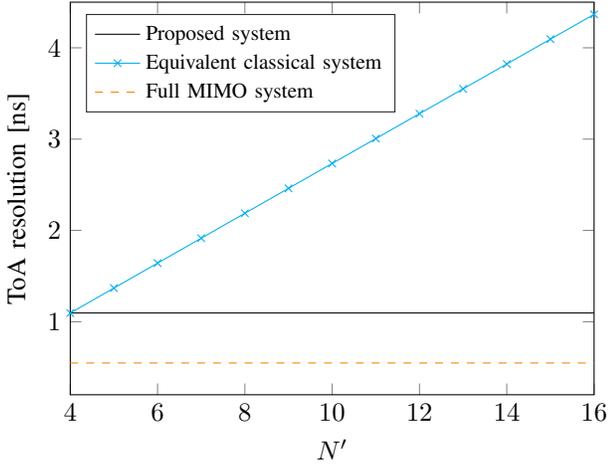
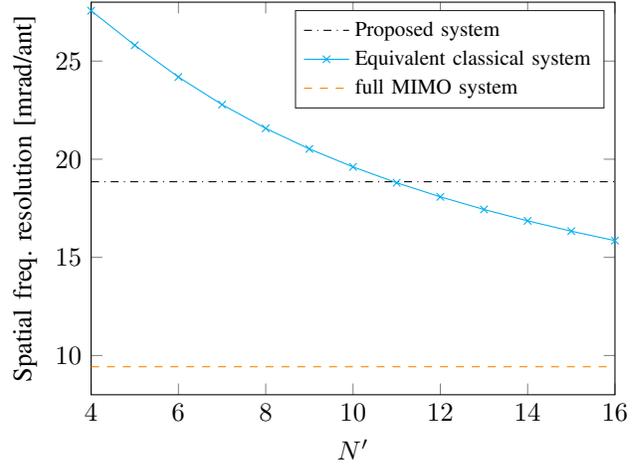

Fig.~\ref{fig:theor_res} illustrates the \gls*{toa} and \gls*{aoa} resolution as a function of the number of antennas for these three systems. In particular, Fig.~\ref{fig:theor_ToA} shows the resolution of the \gls{toa}, whereas in Fig.~\ref{fig:theor_AoA}, we can see the \gls{aoa} resolution in terms of spatial frequency difference (i.e. $\omega_1 - \omega_2$).  We can observe that our proposed architecture has roughly half the resolution of the full \gls*{mimo} architecture in both \gls*{toa} and \gls*{aoa}, while requiring a significantly lower sampling rate. Moreover, it matches or outperforms all the equivalent architectures for \gls*{toa} estimation. It only falls short in \gls*{aoa} estimation with respect to the equivalent classical system for $N'\geq12$ antennas. However, the difference in resolution is up to $15\%$ for $N'=16$ antennas, and it corresponds to a \gls*{toa} resolution that is $4$ times worse. 
This is due to the narrower bandwidth of the equivalent classical system, which allows it to obtain an \gls*{snr} 16 times higher with the same transmitted power. It should be also noted that this analysis applies for individual \gls*{toa} and \gls*{aoa} estimation, whereas the method we propose in the following exploits the \gls*{toa} information to improve the \gls*{aoa} estimation. For this reason, we expect the gap in \gls*{aoa} estimation resolution to be even higher in the final implementation.

Next, we analyzed \gls*{aoa} and \gls*{toa} as two disjoint parameters. However, real applications often rely on joint \gls{aoa} and \gls*{toa} estimation to correctly predict the location of the targets without ambiguity.
While there are computationally heavy methods for joint estimation, such as \gls*{2dmusic}, to the best of our knowledge, there are no known solutions to the problem of matching separately estimated \gls*{aoa} and \gls*{toa} that comes from heterogeneous beamforming architectures. In the following, we will provide a method to overcome this issue, and obtain joint \gls*{toa} and \gls*{aoa} estimates, while still performing the estimation of each parameter independently. This is a key feature to support the rank-bandwidth trade-off that is provided by the proposed architecture, as \gls*{2dmusic} only works with fully digital \gls*{mimo}.

\section{Sensing algorithm} \label{sec:sens}
In this section, we provide an in-depth explanation of the proposed sensing technique, describe its computational complexity and compare it with the complexity of \gls{2dmusic}.

\begin{figure}[t]
\centering
\resizebox{0.95\linewidth}{!}{\begin{tikzpicture}

\node[draw, black, minimum height = 50, minimum width = 80] (toa) at (0, 0) {ToA estimation};

\node[draw, black, minimum height = 50, minimum width = 80, text width = 80, align = center] (amp) at (5, 0) {Path component amplification};

\node[draw, black, minimum height = 50, minimum width = 80, text width = 80, align = center] (aoa) at (10, 0) {AoA estimation};

\draw[-stealth] (-4, 0) -- (toa.west) node[midway,above]{$\hat{H}_A(i,s \Delta_f)$};

\draw[-stealth] (toa.east) -- (amp.west) node[midway,above]{$\left\{\hat{\tau}_\ell \right\}$};

\draw[-stealth] (5,2) -- (amp.north) node[midway, left]{$\hat{H}_m (i,s \Delta_f)$};

\draw[-stealth] (amp.east) -- (aoa.west) node[midway,above]{$\bar{H}_d (\ell,m)$};

\draw[-stealth] (aoa.east) -- (13, 0) node[midway,above]{$\left\{\hat{\theta}_{\ell, q}\right\}$};

\end{tikzpicture}}
\caption{Block diagram of the proposed method}
\label{fig:block}
\end{figure}
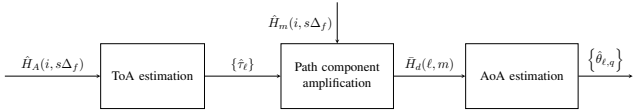

Fig.~\ref{fig:block} illustrates our proposed approach for accurate \gls{toa} and \gls{aoa} estimation. 
First, we estimate the \gls*{toa} of the multipath components by applying a super-resolution algorithm, e.g., \gls{music}, on the \gls*{csi} obtained from the wideband analog beamforming. Given the large bandwidth, the resulting \gls*{toa} estimation is highly accurate. In the second step, for each estimated \gls*{toa}, we process the channel estimates from the narrowband digital beamforming to amplify the component related to the \gls{toa} of interest. In particular, we use \gls*{mrc} over frequency and frame to enhance the component of interest and suppress the other components and the noise. Finally, using matrix pencil, the \gls*{aoa} of each path is estimated from the amplified \gls*{csi}. In the following, these steps are explained in detail.

Without loss of generality, let us assume that we use a set of $F$ consecutive frames starting from frame $0$. The frames are non-coherent, i.e. the phase of each component changes between frames, and $F$ should be chosen such that the targets do not move significantly between the first and last frame (i.e. its position change is much smaller than $\frac{c}{B_A}$).

As described above, the first step to the algorithm is to estimate the \gls{toa} from the analog beamforming channel estimate. While this method is agnostic to the \gls{toa} estimation applied, for clarity, we use the \gls{music} algorithm, described in appendix \ref{sec:music}. 
We assume the \gls{toa} estimation returns a list of peak times $\mathcal{T} = \left\{\hat{\tau}_0,...,\hat{\tau}_{L-1} \right\}$
For each $\hat{\tau}_\ell \in \mathcal{T}$ we estimate the channel coefficient for all frames as 
\begin{equation}
\hat{\alpha}_{(\ell,i)} = \sum_{s = -\left\lceil \frac{S}{2} \right\rceil +1} ^{\left\lfloor \frac{S}{2} \right\rfloor -1} e^{j 2 \pi \tau_\ell s} H_{A}(i, s \Delta_f).
\end{equation}
With this information, we can combine the digital beamforming channel estimates of all frames to amplify a specific component. The digital combined channel responses for  $m \in \mathcal{M}$ and path $\ell$ is computed as:  
\begin{equation}
\bar{H}_d(\hat{\tau}_\ell,m) = \sum_{i=0}^{F-1} \sum_{s=-\left\lceil \frac{S}{2 M} \right\rceil+1}^{\left\lfloor \frac{S}{2M} \right\rfloor -1} \hat{\alpha}_{(\ell,i)}^* e^{j 2 \pi \hat{\tau}_\ell s \Delta_f} \hat{H}_m (i,s \Delta_f). \label{eq:cophasing}
\end{equation}
This is the compensated channel that will be used later for the \gls{aoa} estimation. To show that it indeed this amplifies the component associated with the given \gls{toa}, we replace the channel model in the expression. Assuming a correct \gls{toa} and channel coefficient estimation, i.e. $\hat{\tau}_\ell = \tau_{\hat{k}}$ and $\hat{\alpha}_{(\ell, i)} = \alpha_{(\hat{k},i)}$ for some $\hat{k}$, and a constant beam $\beta_{n,i} = \beta_n$, we obtain\footnote{We omit the limits of $s$ and $i$ for compactness}
\begin{align}
\bar{H}_d & (\hat{\tau}_\ell,m) = \sum_{i, s} \left( \sum_{n=0}^{N-1} \beta_n \sum_{k=0}^{K-1} \alpha_{(\hat{k},i)} e^{j  n \pi \cos(\theta_k) } \delta(  \tau_{\hat{k}} - \tau_k)  \right)^* \cdot  \nonumber  \\
& e^{j 2 \pi \tau_{\hat{k}} s \Delta_f} \left( \sum_{k=0}^{K-1} \alpha_{(k,i)} e^{j  n \pi \cos(\theta_k) } e^{-j 2 \pi s \Delta_f \tau_k} + w \right) .
\end{align}
We now define $ \Xi_n = \sum_{n=0}^{N-1} \beta_n e^{j  n \pi \cos(\theta_{\hat{k}}) }$ for compactness. We also note that the noise statistic is unaffected by a rotation in the complex plane, hence we can rewrite the expression as
\begin{align} \label{eq:mrc_final}
\bar{H}_d(\hat{\tau}_\ell,m) = \Xi_n \sum_{i, s}  \alpha^*_{(\hat{k},i)}  \sum_{k=0}^{K-1} & \alpha_{(k,i)} e^{j n \pi \cos(\theta_k) } \cdot \nonumber \\& e^{j 2 \pi s \Delta_f (\tau_{\hat{k}} - \tau_k)} + w.
\end{align}
Without loss of generality, we assume $\hat{k} = 0$ and $ \Xi_n = 0$. Then Eq. \eqref{eq:mrc_final} can expressed as:
\begin{align}
\bar{H}_d&(\hat{\tau}_\ell,m) = \sum_{i, s} \vert \alpha_{(1,i)}\vert^2 e^{j n \pi \cos(\theta_1) } \\
& +\sum_{k=1}^{K-1} \sum_{i, s}     \alpha^*_{(\hat{k},i)} \alpha_{(k,i)} e^{j  n \pi \cos(\theta_k) } e^{j 2 \pi s \Delta_f (\tau_{\hat{k}} - \tau_k)} \\
& +\sum_{i, s}    \alpha^*_{(\hat{k},i)}  w.
\end{align}
Here we can see that the component with $k=0$ is combined coherently, whereas the other components, as well as the noise, are combined incoherently. This ensures that the component of interest is amplified. 
This operation is the core novelty of the proposed method, as it allows the \gls{toa} and \gls{aoa} association that would be otherwise impossible because the narrowband part does not have enough bandwidth. The key observation here is that while the wideband analog and narrowband digital domains do not share a common \gls{aoa} or \gls{toa} information that can be reliably used for the association, they do share the doppler domain, so the phase change over multiple frames is the only possible mean to connect the \gls{toa} and \gls{aoa} results.

After computing the digital combined channel for each path, we perform \gls{aoa} estimation on each of the digitally combined channels. Again, the proposed scheme can work with different \gls{aoa} estimation methods. In this case though, we note that the combination produces a single vector, therefore methods that relies on the singular value decomposition like \gls{music} are not suitable to detect more than one angle. For this reason, we use the Matrix Pencil method, described in appendix \ref{sec:MP}.

We apply the such method to the vector $\bar{\bf{H}}_d(\hat{\tau}_\ell) = \left( \bar{H}_d(\hat{\tau}_\ell,0), ..., \bar{H}_d(\hat{\tau}_\ell,M-1) \right)$ with a pencil parameter $P$ to obtain the estimated spatial frequencies and amplitude of the components. For the generic $\ell^{\textit{th}}$ component, we obtain a set of spatial frequencies $ \bf{\Omega}_\ell = \left\{ \omega_{\ell,0}, ..., \omega_{\ell, M-P-1}\right\}$. For each spatial frequency, thus obtaining the set of amplitudes $\mathcal{A}_\ell = \left\{ \hat{a}_{\ell,0}, ..., \hat{a}_{\ell, M-P-1}\right\}$, where:
\begin{equation}
\hat{a}_{\ell,q} = \sum_{d=0}^{M-1}  \bar{H}_d(\hat{\tau}_\ell,q) e^{-j \frac{2 \pi d}{M} \omega_{\ell,q}}.
\end{equation}
Assuming that the number of components associated with the time domain sample $\hat{\tau}_\ell$ are less than $M-P$, some of the components reported will be a result of noise. Therefore, we apply a threshold to the amplitudes $\mathcal{A}$ to remove such components. In particular, the threshold is defined as:
\begin{equation}
\textrm{Thr} = \rho \sqrt{ \frac{1}{M} \sum_{m=0}^{M-1} \vert \bar{H}_d(\hat{\tau}_\ell,m) \vert^2 }, \label{eq:thr}
\end{equation}
where $\rho$ is a configurable parameter.\footnote{Minimum Description Length Criterion or Akaike Information Criterion could also be used.} 
The components $\left\{q : \: \vert \hat{a}_{\ell,q} \vert > \textrm{Thr} \right\}$ are kept, and their angle is estimated as $\hat{\theta}_{\ell, q} = cos^{-1} \left( \frac{2 \omega_{\ell,q}}{M} \right)$. This will generate the set $\mathcal{Z} = \{Z_1,...,Z_O\}$ of output range-angle pairs of the form $Z_o = \left(\hat{\tau}_\ell, \hat{\theta}_{\ell, q}\right)$. For the sake of a lighter notation, we define $\tau(Z_o) = \hat{\tau}_\ell$ and  $\theta(Z_o) =  \hat{\theta}_{\ell, q}$. The complete procedure is summarized in Algorithm \ref{alg:sens}.

\begin{algorithm}
\caption{Sensing algorithm}
    \begin{algorithmic}[1]
    \State $\mathcal{Z} \gets \{\}$
    \State $\mathcal{T} \gets MUSIC(\hat{H}_A(i,s))$
    \ForEach {$\ell \in \{0,...,L-1\}$}        
    \State $\hat{\alpha}_{\ell,i} \gets \sum_{s = -\left\lceil \frac{S}{2} \right\rceil+1} ^{\left\lfloor \frac{S}{2} \right\rfloor-1} e^{j 2 \pi \hat{\tau}_\ell s} H_{A}(i, s \Delta_f)$
    \State $\bar{H}_d(\ell,m)  \gets \sum_{i,s} \hat{\alpha}_{(\ell,i)}^* e^{j 2 \pi \hat{\tau}_\ell s \Delta_f} \hat{H}_m (i,s \Delta_f)$
    \State $\textrm{Thr} \gets \rho \sqrt{ \frac{1}{M} \sum_{m=0}^{M-1} \vert \bar{H}_d(\hat{\tau}_\ell,m) \vert^2 }$
    \State $\bf{\Omega}_\ell \gets MatrixPencil(\bar{\bf{H}}_d(\hat{\tau}_\ell))$
    \ForEach {$\omega_{\ell,q} \in \bf{\Omega}_\ell$}   
        \State $\hat{a}_{\ell,q}  \gets \sum_{d=1}^{M}  \bar{H}_d(\hat{\tau}_\ell,q) e^{-j \frac{2 \pi d}{M} \omega_{\ell,q}}$
        \If{$\vert \hat{a}_{\ell,q} \vert \geq \textrm{Thr}$}
            \State $\mathcal{Z} \gets  \mathcal{Z} \cup \left\{ \tau_\ell,  cos^{-1} \left( \frac{2 \omega_{\ell,q}}{M} \right) \right\}$
        \EndIf
    \EndFor
    \EndFor
    \end{algorithmic}  
\label{alg:sens}
\end{algorithm}

\begin{figure*}[ht!]
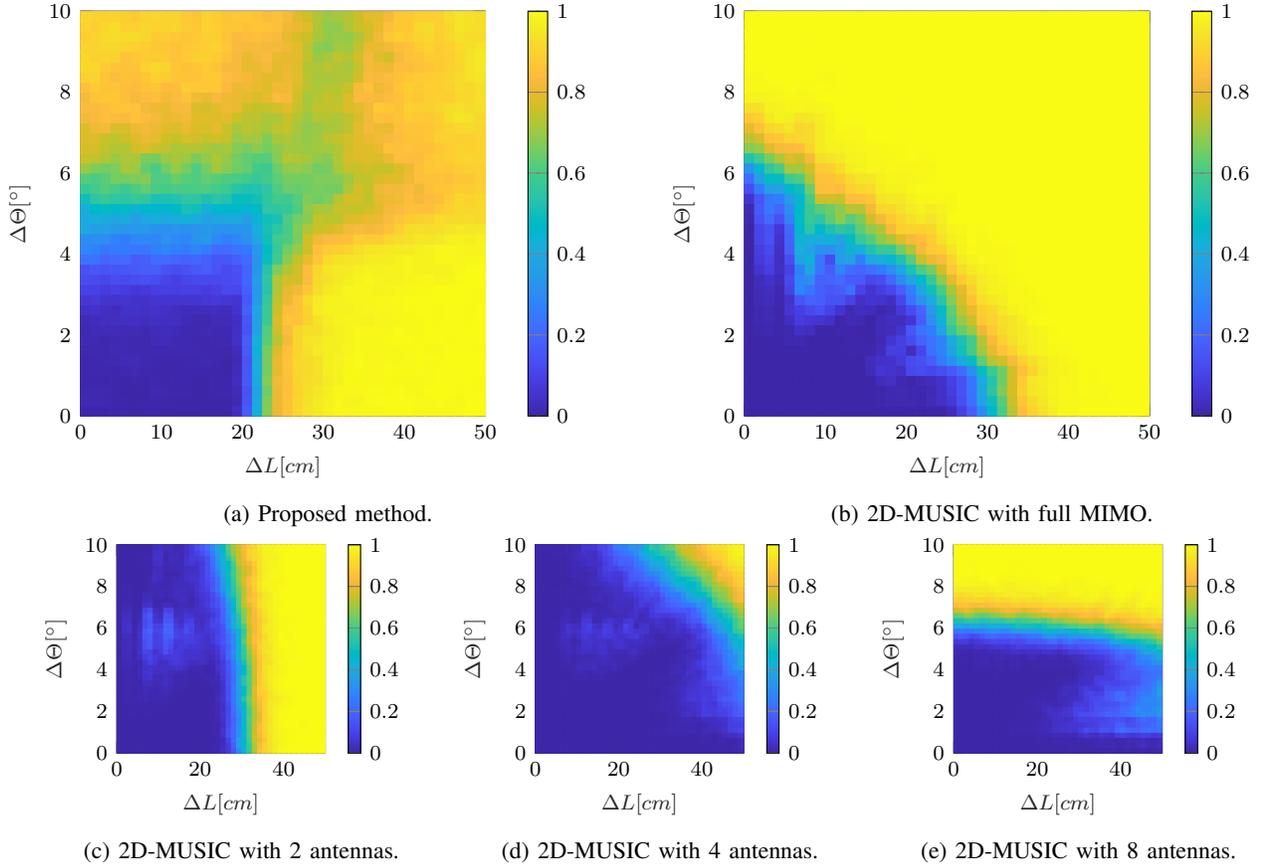

     \centering
     \begin{subfigure}[t]{0.48 \linewidth}
         \input{figures/resolution}
         \caption{Proposed method.}
         \label{fig:res_coph}
     \end{subfigure}
     \begin{subfigure}[t]{0.48 \linewidth} 
         \input{figures/resolution_full}
         \caption{\gls{2dmusic} with full \gls{mimo}.}
         \label{fig:res_2dm}
     \end{subfigure}     
     \begin{subfigure}[t]{0.3 \linewidth}
         \input{figures/resolution_2ants}
         \caption{\gls{2dmusic} with 2 antennas.}
         \label{fig:res_4ant}
     \end{subfigure}
     \begin{subfigure}[t]{0.3 \linewidth}       
         \input{figures/resolution_4ants}
         \caption{\gls{2dmusic} with 4 antennas.}
         \label{fig:res_8ant}
     \end{subfigure}
     \begin{subfigure}[t]{0.3 \linewidth}       
         \input{figures/resolution_8ants}
         \caption{\gls{2dmusic} with 8 antennas.}
         \label{fig:res_16ant}
     \end{subfigure}
        \caption{Resolution probability of the proposed method, showing similar performance to full bandwidth fully digital \gls{mimo} and widely outperforming \gls{mimo} systems with equivalent sampling rate.  }
        \label{fig:resolution}
\end{figure*}

\subsection{Computational complexity}
One of the main advantages of our proposed method is its low computational complexity. The improvement in performance is mainly due to breaking down the problem into two different 1-D estimation stages, as opposed to the joint 2D estimation used in, e.g., \gls{2dmusic}, thus removing the need to search in a 2D parameter space.  
In particular, if we consider 2-D \gls{music}, the computational complexity\footnote{assuming the execution time is dominated by the eigenvalue decomposition and  music spectrum computation} is $\mathcal{O}((SN)^3 + (SN)^2n_{\theta}n_{\tau})$, where $n_{\theta}$ and $n_{\tau}$ are the sizes of the search grid in angle and range, respectively. By contrast, our method reduces the complexity to $\mathcal{O}(S^3 + S^2 n_{\theta} + K F S + K N^3)$. Where the terms of the summation refer respectively to the matrix decomposition for the 1-D \gls{music} ($S^3$), the spectrum computation for the 1-D \gls{music} ($S^2 n_{\tau}$), the $K$ path component amplification steps ($K F S$) and the execution of Matrix Pencil ($K N^3$). It should be noted that the matrix pencil contribution only depends on the number of antennas and components, which are typically much smaller than the number of subcarriers. Therefore it often has a negligible contribution and can be ignored in most practical situations. Moreover, we note that the complexity of the path component amplification is not dependent on the number of antennas used since using more antennas results in a lower number of subcarriers for the digital beamforming. In order to better understand the difference in performance, we study the effect of scaling all the parameters (i.e. $S$, $N$, $n_{\tau}$, $n_\theta$ and $K$) by a factor $\alpha$. In the case of \gls{2dmusic} the complexity becomes $\mathcal{O}(\alpha^6 \left[(SN)^3 + (SN)^2n_{\theta}n_{\tau} \right])$, whereas the proposed method has complexity $\mathcal{O}(\alpha^3 \left[S^3 + S^2 n_{\tau} + K F S + K N^3 \right])$. For example, if $\alpha=2$, i.e., we double all the parameters in the system, the proposed method will take $8$ times longer to execute, whereas \gls{2dmusic} will take $64$ times longer. Indeed, from this example we can understand how the proposed method is much more scalable.

\section{Results} \label{sec:results}

\subsection{Numerical analysis} \label{sec:numres}
For this analysis, we generate a channel according to the model described in \ref{sec:sysmod}, with a bandwidth $B_A$ = 400~MHz, $S$ = 128 subcarriers, $F$ = 10 frames and an \gls*{snr} of 10~dB on each antenna. We use a 2 paths channel with $\theta_0 = 15^\circ$, $\tau_0 = \frac{10\:\text{m}}{c}$, and  $\theta_1 = 30^\circ + \Delta \theta$ $\tau_1 = \frac{10\:\text{m} + \Delta L}{c}$, respectively. The complex amplitudes $\alpha_0$ and $\alpha_1$ have the same statistics, corresponding to a complex normal distribution of zero mean and the identity as covariance matrix. Moreover, we use $\rho = 0.3$ for the threshold in Eq.~\eqref{eq:thr}.  We run the algorithm for $(\Delta \theta, \Delta L) \in \left\{ 0\:\text{m}, 0.0125\:\text{m}, \ldots, 0.5\:\text{m}   \right\} \times \left\{0^\circ, 0.25^\circ,\ldots,10^\circ  \right\}$. For each  $(\Delta \theta, \Delta L)$ pair, we generate $100$ realizations of the complex channel coefficients, as well as the noise. For each realization, we compare the output $\{Z_1,\ldots,Z_O\}$ with the real channel parameters as followings.
For each estimated component $Z_o$, we find the closest true component according to the distance function
\begin{equation}
d\left( Z_o, (\tau_k, \theta_k) \right) = \sqrt{\frac{1}{\sigma^2_\tau} (\tau(Z_o) - \tau_k)^2 +\frac{1}{\sigma^2_\theta} (\theta(Z_o) - \theta_k)^2 }.
\end{equation} 
Without loss of generality, let us assume that the closest component is the first one. In case $d\left( Z_o, (\tau_0, \theta_0) \right) < 1$, we associate the estimated component $Z_o$ with the first component, otherwise it is considered a spurious component. If both real components have an associated estimated component, we say that the components have been resolved. Moreover, the said component is used to determine the \gls*{rms} error for the angle and delay estimation. For this evaluation, we use $\sigma_\tau = \frac{30\:\text{cm}}{c}$ and $\sigma_\Theta = 3^\circ$.

The resulting resolution probability for the proposed method, as well as 2-D \gls*{music} on the full \gls*{mimo} system is shown in Fig.~\ref{fig:res_coph} and \ref{fig:res_2dm}, respectively. It should be noted that for \gls{2dmusic} the correct number of components, and thus the size of the noise subspace, is {\em assumed to be know exactly} and is provided as an input of the algorithm. In a real deployment, this parameter would need to be estimated as well, potentially further reducing the resolution capabilities of the method.
We can see that the proposed method, despite using only $25\%$ the total sampling rate and having much lower complexity, performs comparably to the classical \gls{2dmusic} in terms of resolution. At least for the time domain, this is not unexpected: According to Eq.~\eqref{eq:toa_res} in fact, the dependence on antenna and \gls{snr} of the resolution is proportional to $ \sqrt{4}{\frac{1}{\Gamma N}}$. Despite having $N=1$ in the proposed architecture, we have that the \gls{snr} is $\Gamma = \Gamma_0 N$ thanks to the beamforming, so the product $\Gamma N$ stays constant. In Fig.~\ref{fig:res_4ant}, \ref{fig:res_8ant} and \ref{fig:res_16ant} we can observe the resolution for \gls{2dmusic} for a digital beamforming system with the same aggregate sampling rate  as the proposed architecture. Here, the sampling rate is spent for either digitalizing multiple antennas or a larger bandwidth. We may easily notice that the proposed method exhibits a far superior resolution to \gls{2dmusic}, either in range, angle or both, based on the allocation of the sampling rate.

\begin{figure}[t]
\centering
\input{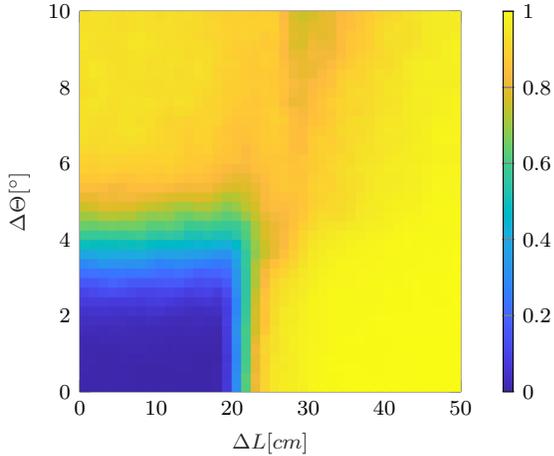}
\caption{Resolution probability of the proposed method with a fully digital MIMO architecture, showing similar performance to the proposed architecture while requiring a much larger aggregate sampling rate.}
\label{fig:newarch_full}
\end{figure}

Finally, In Fig.~\ref{fig:newarch_full}, we can see the resolution achieved by the proposed method when provided with digital beamforming for the full bandwidth. The result is very close to the one show in Fig.~\ref{fig:res_coph}. This clearly show that, at least from a resolution standpoint, fully digital \gls{mimo} on the whole bandwidth is not providing a huge advantage, and certainly does not justify the huge increase in cost and power consumption.

\begin{figure}[t]
     \centering
     \begin{subfigure}[t]{0.8 \linewidth}
     \centering
%
%

\begin{tikzpicture}

\begin{axis}[%
width=\fwidth,
height=\fheight,
scale only axis,
xmin=0,
xmax=15,
xlabel style={font=\color{white!15!black}},
xlabel={$\Gamma_0$},
ymin=0,
ymax=9,
xlabel near ticks,
ylabel style={font=\color{white!15!black}},
ylabel={Range RMSE [mm]},
ylabel near ticks,
legend columns = 4,
axis background/.style={fill=white},
legend style={legend cell align=left, align=left, draw=white!15!black, at={(0.5,1.02)}, anchor=south, font=\footnotesize, style={column sep=0.25cm}}
]

\addplot [color=mycolor3]
  table[row sep=crcr]{%
0	5.41717459663713\\
3	4.31590954531276\\
6	3.03193900449511\\
9	2.11268091934828\\
12	1.53095341725155\\
15	1.1368710953413\\
};
\addlegendentry{N'=2}

\addplot [color=mycolor4]
  table[row sep=crcr]{%
0	6.55829893686929\\
3	5.34425088634925\\
6	4.11224294077991\\
9	3.11861548449878\\
12	2.1960732634037\\
15	1.54945205152244\\
};
\addlegendentry{N'=4}

\addplot [color=mycolor5]
  table[row sep=crcr]{%
0	7.60161262545954\\
3	6.68425602283097\\
6	5.47695316424384\\
9	4.11584367482052\\
12	2.98578508950619\\
15	2.12095567852584\\
};
\addlegendentry{N'=8}

    \addlegendimage{empty legend}
    \addlegendentry{}

\addplot [color=mycolor1, mark=x, mark options={solid, mycolor1}]
  table[row sep=crcr]{%
0	7.33816437029424\\
3	4.71356818955596\\
6	2.60004894995553\\
9	1.77666553864637\\
12	0.916797246797988\\
15	0.644623799957102\\
};
\addlegendentry{\makebox[5pt][l]{Proposed architecture}}

    \addlegendimage{empty legend}
    \addlegendentry{}

\addplot [color=mycolor2, mark=o, mark options={solid, mycolor2}]
  table[row sep=crcr]{%
0	3.00908487642705\\
3	2.13118669008681\\
6	1.51632549326928\\
9	1.10751852511977\\
12	0.818987462619604\\
15	0.607805806332811\\
};
\addlegendentry{\makebox[5pt][l]{Full MIMO}}

    \addlegendimage{empty legend}
    \addlegendentry{}

\end{axis}

\end{tikzpicture}%
         \caption{Range \gls{rmse}.}
         \label{fig:accuracy_vs_rms_range}
     \end{subfigure}
     \begin{subfigure}[t]{0.8 \linewidth}  
     \centering
%

\begin{tikzpicture}

\begin{axis}[%
width=\fwidth,
height=\fheight,
scale only axis,
xmin=0,
xmax=15,
xlabel style={font=\color{white!15!black}},
xlabel={$\Gamma_0$},
xlabel near ticks,
ymin=0,
ymax=0.5,
ylabel style={font=\color{white!15!black}},
ylabel={Angle RMSE [$^{\circ}$]},
ylabel near ticks,
axis background/.style={fill=white},
]
\addplot [color=mycolor1, mark=x, mark options={solid, mycolor1}, forget plot]
  table[row sep=crcr]{%
0	0.150144153995307\\
3	0.114250436894903\\
6	0.0843760323354754\\
9	0.0608086247143401\\
12	0.0548424717231764\\
15	0.0349166732236069\\
};

\addplot [color=mycolor2, mark=o, mark options={solid, mycolor2}, forget plot]
  table[row sep=crcr]{%
0	0.0615549290963424\\
3	0.0447929397433339\\
6	0.0312952810890324\\
9	0.0224074106091981\\
12	0.0154528747385936\\
15	0.0110257198942263\\
};

\addplot [color=mycolor3, forget plot]
  table[row sep=crcr]{%
0	0.509667949495227\\
3	0.387492157908981\\
6	0.283384151401951\\
9	0.210564287153655\\
12	0.139105576859013\\
15	0.0968096319076648\\
};

\addplot [color=mycolor4, forget plot]
  table[row sep=crcr]{%
0	0.177622824615653\\
3	0.129129771917569\\
6	0.0877528658579927\\
9	0.061486633359018\\
12	0.042475523481899\\
15	0.031573744072455\\
};

\addplot [color=mycolor5, forget plot]
  table[row sep=crcr]{%
0	0.0594837238652504\\
3	0.0437409627977439\\
6	0.0307307862870261\\
9	0.0211700596496629\\
12	0.0156465216970983\\
15	0.0111331671093306\\
};

\end{axis}
\end{tikzpicture}%
         \caption{Angle \gls{rmse}.}
         \label{fig:accuracy_vs_rms_angle}
     \end{subfigure}
        \caption{Accuracy of the proposed method compared to \gls{2dmusic}, showing comparable performance dispite the much lower aggregate sampling rate.}
        \label{fig:accuracy_vs_rms}
\end{figure}
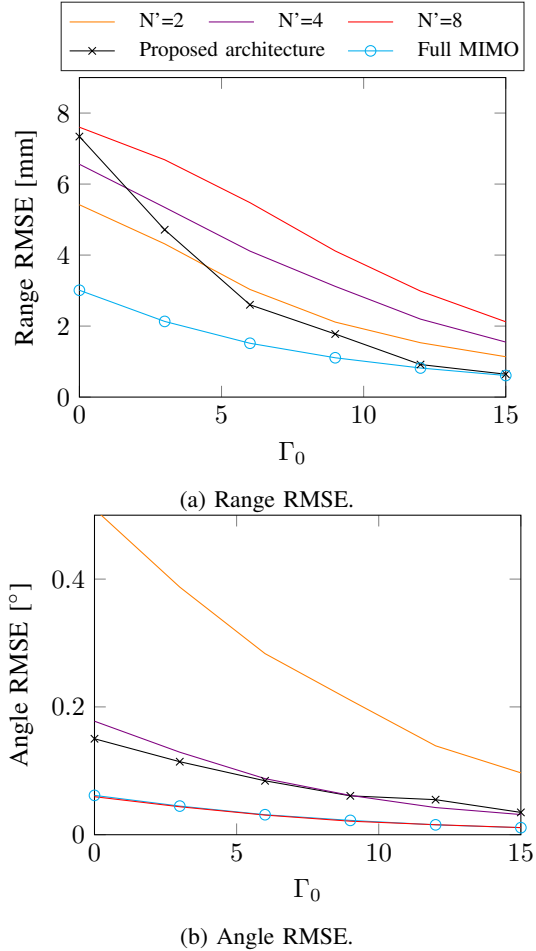

In Fig.~\ref{fig:accuracy_vs_rms} we show the accuracy of the parameters estimated by the proposed method, as well as the one estimated by \gls{2dmusic} for the full \gls{mimo} system and for the equivalent \gls{mimo} systems with $N' = \{2, 4, 8\}$ antennas. We generated a channel composed of a single component with a random angle and range, and ran the algorithm on a such channel to evaluate the error. We repeated the process for $1000$ channel realizations to obtain the estimated \gls{rmse}, which is plotted as a function of the full bandwidth \gls{snr} $\Gamma_0$. The actual \gls{snr}, however, has been adjusted for the equivalent \gls{mimo} systems to $\gamma = \gamma_0 \frac{2}{N'}$.

As it can be seen in the figure, at high \gls{snr} the performance of the proposed method is comparable with the fully-digital \gls{mimo} system, while they slightly degrade towards the low \gls{snr} regime. This degradation is likely due to the fact that in the simulation we use only one frame to design the beamforming coefficients, thus, the beam is affected by noise.

In terms of angle, the system with $N'=8$ antennas seems to largely outperform the proposed method. It should be noted however that this advantage is caused to the higher SNR due to the lower total bandwidth, which also implies a large degradation in terms of communication due to the low bandwidth.

\section{Discussion} \label{sec:discussion}

\subsection{Leakage issue}
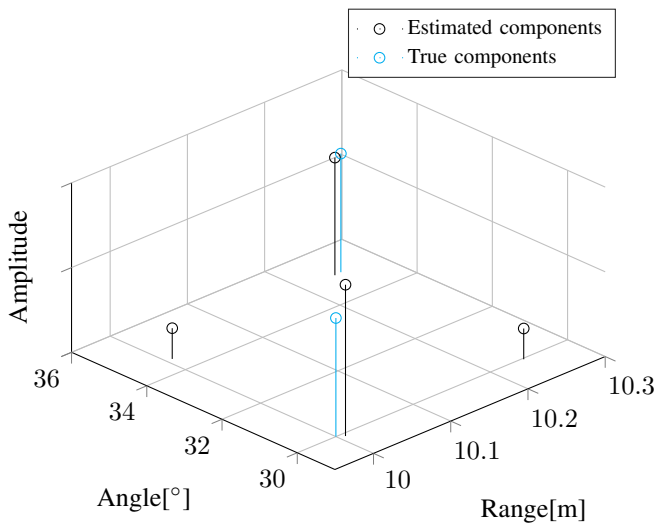
\begin{figure}[t]
\centering
%
%

%
\begin{tikzpicture}

\begin{axis}[%
width=\fwidth,
height=\fheight,
scale only axis,
xmin=9.95,
xmax=10.3,
tick align=outside,
ymin=29,
ymax=36,
zmin=0,
zmax=10000000,
view={-44.099999700588}{43.7999996350061},
axis background/.style={fill=white},
axis x line*=bottom,
axis y line*=left,
axis z line*=left,
xmajorgrids,
ymajorgrids,
zmajorgrids,
ylabel={Angle[$^\circ$]},
xlabel={Range[m]},
zlabel={Amplitude},
scaled ticks=false,
zticklabels ={,,},
legend style={legend cell align=left, align=left, draw=white!15!black, at={(1.02,0.98)}, anchor=south east, font=\footnotesize}
]
\addplot3 [ycomb, color=mycolor1, mark=o, mark options={solid, mycolor1}]
 table[row sep=crcr] {%
10.0072766519886	34.4996925954909	1814028.99928655\\
10.0072766519886	29.8938833894739	8931504.64768812\\
10.2414895098011	34.9837101670488	6941472.93595117\\
10.2414895098011	29.9645948741868	1815216.51919981\\
};
\addlegendentry{Estimated components};
 \addplot3 [ycomb, color=mycolor2, mark=o, mark options={solid, mycolor2}]
 table[row sep=crcr] {%
10	30	7000000\\
10.25	35	7000000\\
};
\addlegendentry{True components};

 \end{axis}

\end{tikzpicture}%
\caption{Example of leakage.}
\label{fig:leak}
\end{figure}

Due to imperfections in the coefficient estimation, when two targets are very close we often observe that some of the components are also visible at different times. An example is shown in Fig.~\ref{fig:leak}, where we may notice the algorithm estimation result in the case where the channel has two components (depicted in orange) at $(10~\text{m}, 30^\circ)$ and $(10.25~\text{m}, 35^\circ)$. As we can see, the two components are correctly identified, but the algorithm also detects two leaked components at  $(10~\text{m}, 35^\circ)$ and $(10.25~\text{m}, 30^\circ)$. When the distances get closer, these components may have a significant amplitude, and sometimes they are not filtered out with a simple threshold. Moreover, if the amplitude of one component is significantly larger than the others, we may even observe that at a specific time, the real component has a lower amplitude than the leaked one. This unfortunately means that it is possible to misidentify some components when using the simple threshold strategy proposed in \ref{sec:numres}. Over the whole resolution experiment, we recorded an overall probability of observing at least a leaked component of $25\%$, i.e., we reported at least one spurious component over at least one-quarter of the channel realizations. On average, we observed $0.41$ leaked components per realization. This suggests that the leakage problem is significant, at least for close targets. It should be noted though, that for more spread targets the issue is a lot less relevant. For example, repeating the expertiment with $(\Delta \theta, \Delta L) \in \left\{ 0\:\text{m}, 1\:\text{m}, \ldots, 5\:\text{m}   \right\} \times \left\{0^\circ, 1^\circ,\ldots,10^\circ  \right\}$, the average fraction of observations showing leakage reduces to $0.98\%$, with an average number of leaked component of $0.017$. Moreover, it is always true that a real component is larger than the leaked component with the same angle, so it is possible to replace the threshold a better classification algorithm to identify the real components after the estimation. However, we will address this solution in future work.

\subsection{Target illumination}

\begin{figure}[t]
\centering
\input{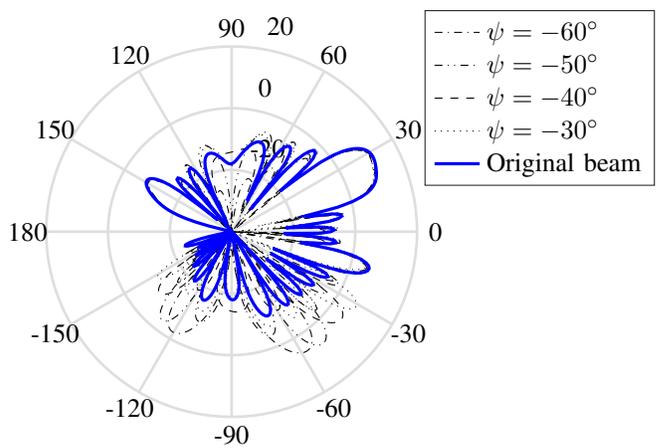}
\caption{Beamformed farfield pattern [dBi] without the additional lobe (blue) and with an additional lobe in different directions (black)}
\label{fig:pattern}
\end{figure}

Naturally, in order to detect an object that should be both illuminated by the transmitter and fall within the receiver beam.
In order to verify if this is the case, we can compute the \gls*{af} of the \gls*{mrc} beam, which is 
\begin{align}
\text{AF}(\phi) &= \sum_{n=0}^{N-1} \left(\sum_{k=0}^{K-1} \alpha_k e^{j n \pi \cos(\theta_k)} \right)^*  e^{j n \pi \cos(\phi)}\\
&=  \sum_{k=0}^{K-1} \alpha^*_k \sum_{n=0}^{N-1} e^{-j  n  \pi \cos(\theta_k)} e^{j  n \pi \cos(\phi)}.
\end{align}
This shows that the \gls*{af} generated by the \gls*{mrc} can be decomposed in a linear combination of beams with expression 
\begin{equation}
\text{AF}_k(\phi) = \sum_{n=0}^{N-1} e^{-j  n \pi \cos(\theta_k)} e^{j n \pi \cos(\phi)}.
\end{equation}
Notably, $\text{AF}_k(\phi)$ is the array factor of a beam pointed in the direction $\theta_k$. If \gls*{mrc} is performed on both ends, this fact should guarantee both illumination of the target and a suitable receiver beam.\\
Moreover, to improve the \gls*{snr} of low amplitude components, we propose to add a perturbation to the transmitter beam. We define the new beamforming coefficients as \begin{equation}
\beta'_n(\xi, \varphi) = \xi e^{j  n \pi \cos(\varphi)} + (1 - \xi) \beta_n.
\end{equation}
Where $\beta_n$ are the beamforming coefficients derived by the \gls*{mrc}, $\xi \in (0,1)$. This adds a lobe in the direction $\varphi$ with an amplitude $\xi$, which could potentially illuminate better some far or low reflective objects. An example of the resulting array factor farfield pattern with and without the additional lobe can be seen in Fig.~\ref{fig:pattern}, where we show in blue the original pattern, and in dashed black we can see 4 examples with an additional side lobe at $-60^\circ$, $-50^\circ$, $-40^\circ$ and $-30^\circ$ respectively. The sidelobe has been added with a relative amplitude of $\xi = 0.2$. As it can be clearly seen, the addition of the side lobe has a negligible impact on the gain of the main pattern.

 It should be noted that better illumination of those objects will cause their component at the receiver to have a higher amplitude, and thus to be amplified even further by the receiver \gls*{mrc}. The direction $\varphi$ can then be swept to make sure that every possible object is illuminated. Moreover, since we use only a fraction of the energy equal to $(1-\xi)^2$ in that beam, the communication should receive a penalty in \gls*{snr} of roughly $-20 \log(1-\xi)$, which for small $\xi$ values should be negligible (e.g., $\xi = 0.2$ will cause a loss of roughly $2$dB). Moreover, \gls{mrc} generates a beam that is suitable for communication, as the presence of major lobes in the direction of the received power ensures a high \gls{snr}.

\section{Conclusions} \label{sec:conclusions}
In this paper, we described a low-complexity method to extract \gls{toa} and \gls{aoa} information from a novel hardware architecture that features a combination of a high-bandwidth analog beamforming and low-bandwidth digital beamforming system. We have shown how it is possible to match the \gls{toa} estimate to the relevant \gls{aoa} estimate by acquiring multiple non-coherent frames and using the phase difference between different \gls{toa} to isolate a specific component in the narrowband digital beamforming domain. Finally, we have conducted a numerical evaluation of the performance of the method, showing that it matches or outperforms \gls*{2dmusic} despite the use of significantly fewer \gls*{adc} samples. Finally, the proposed architecture also allows for the design of a robust communication beam by extracting the channel coefficients from the digital beamforming part and applying \gls{mrc}. Such a beam turns out to be also well suited for sensing as it guarantees to capture energy from all possible directions.

\appendices
\section{MUSIC} \label{sec:music}
\gls{music} is a subspace-based super-resolution algorithm that relies on the eigenvalue decomposition of the sample covariance matrix of the received signal as follows:
\begin{equation}
\boldsymbol{R}_y = \boldsymbol{Y} \boldsymbol{Y}^H
\end{equation}
where $\boldsymbol{Y}$ is computed as:
\begin{equation}
\boldsymbol{Y} = \left[ \begin{matrix}
\hat{H}_A \left(0, -\left\lceil \frac{S}{2} \right\rceil +1 \right) & \cdots & \hat{H}_A \left(F-1, -\left\lceil \frac{S}{2} \right\rceil +1 \right)\\
 \vdots & & \vdots \\
 \hat{H}_A \left(0, \left\lfloor \frac{S}{2} \right\rfloor -1\right) & \cdots &  \hat{H}_A \left(F-1, \left\lfloor \frac{S}{2} \right\rfloor -1 \right)
\end{matrix}\right].
\end{equation}
The covariance matrix $\boldsymbol{R}_y$ can be re-written as 
\begin{equation}
\boldsymbol{R}_y=\boldsymbol{A}\boldsymbol{R}_s\boldsymbol{A}^H+\sigma^2_n\boldsymbol{I} \label{eq:Ry}
\end{equation} 
where $\boldsymbol{A}$ and $\boldsymbol{R}_s$ are as defined in \eqref{eq:A} and \eqref{eq:Rs} respectively.

\begin{figure*}[h!]

\begin{equation}
\boldsymbol{A} = \left[ \begin{matrix}
e^{-j 2 \pi \left( -\left\lceil \frac{S}{2} \right\rceil +1 \right) \Delta_f \tau_0} & \cdots & e^{-j 2 \pi \left( -\left\lceil \frac{S}{2} \right\rceil +1 \right) \Delta_f \tau_{K-1}}  \\
 \vdots & & \vdots \\
 e^{-j 2 \pi \left(\left\lceil \frac{S}{2} \right\rceil  -1 \right) \Delta_f \tau_0}  & \cdots & e^{-j 2 \pi \left( \left\lceil \frac{S}{2} \right\rceil -1 \right) \Delta_f \tau_{K-1}}  
\end{matrix}\right]. \label{eq:A}
\end{equation}
\begin{equation}
\boldsymbol{R}_s = \boldsymbol{\chi} \boldsymbol{\chi}^H, \:
\boldsymbol{\chi} = \left[ \begin{matrix}
\sum_{n=0}^{N-1} \beta_{n,i} \alpha_{(0,0)} e^{j  n \pi \cos(\theta_1) }  & \cdots & \sum_{n=0}^{N-1} \beta_{n,i} \alpha_{(0,F-1)} e^{j  n \pi \cos(\theta_1) }  \\
 \vdots & & \vdots \\
\sum_{n=0}^{N-1} \beta_{n,i} \alpha_{(K-1,0)} e^{j  n \pi \cos(\theta_K) } & \cdots & \sum_{n=0}^{N-1} \beta_{n,i} \alpha_{(K-1,F-1)} e^{j  n \pi \cos(\theta_K) } 
\end{matrix}\right]. \label{eq:Rs}
\end{equation}
\hrule
\end{figure*}

$\sigma^2_n$ is the noise power, and $\boldsymbol{I}$ is the identity matrix. The intuition behind the method comes from this decomposition: we can notice that the image of $R_y$ can be decomposed into two subspaces:
\begin{itemize}
\item The signal space, associated with the first term of Eq. \eqref{eq:Ry}, which is the subspace generated by the columns of $\boldsymbol{A}$.
\item The noise subspace, associated with the second term of Eq. \eqref{eq:Ry}, which is the subspace orthogonal to the signal subspace. 
\end{itemize}
We can also infer that, given a sufficient \gls{snr}, the eigenvalues associated with the signal subspace will be significantly larger than the ones associated with the noise subspace.
\gls*{music} extracts the noise subspace by removing the eigenvectors corresponding to the largest eigenvalues of $\boldsymbol{R}_y$ as the vectors that span the signal subspace. The remaining eigenvectors, which correspond to the near-zero eigenvalues, constitute the noise matrix $\boldsymbol{U}_n$. Finally, we can compute the \gls*{music} spectrum, $P_\text{MUSIC}(\tau)=\frac{1}{\boldsymbol{a}^H(\tau)\boldsymbol{U}_n \boldsymbol{U}^H_n \boldsymbol{a}(\tau)}$, where $\boldsymbol{a}(\tau)=[1, e^{-j2\pi \Delta_f  \tau},...,e^{-j2\pi (S-1) \Delta_f  \tau}]^T$ is the steering vector associated with the delay $\tau$. \\
By definition of the noise subspace, if $\tau = \tau_k$ for some $k$, due to the subspace orthogonality, $\boldsymbol{a}^H(\tau)\boldsymbol{U}_n \boldsymbol{U}^H_n \boldsymbol{a}(\tau) \approx 0$, and therefore the \gls{music} spectrum has a large value.
Therefore, we detect this condition by performing a peak detection on $P_\text{MUSIC}(\tau)$ and report the list of peak times $\mathcal{T} = \left\{\hat{\tau}_0,...,\hat{\tau}_{L-1} \right\}$

\section{Matrix Pencil}  \label{sec:MP}
The Matrix Pencil method works as follows.
From the vector $\bar{\bf{H}}_d(\hat{\tau}_\ell)$, we generate a Hankel matrix:
\begin{equation}
\mathcal{H}_{\ell, P} = \left[ \begin{matrix}
\bar{H}_d(\hat{\tau}_\ell,0) & ... & \bar{H}_d(\hat{\tau}_\ell,P-1)\\
\bar{H}_d(\hat{\tau}_\ell,1) & ... & \bar{H}_d(\hat{\tau}_\ell,P)\\
\vdots & \vdots & \vdots \\
\bar{H}_d(\hat{\tau}_\ell,M-P-1) & ... & \bar{H}_d(\hat{\tau}_\ell,M-1)\\
\end{matrix}\right].
\end{equation} 
From this matrix, we generate the two matrices $\mathcal{H}^{(1)}_{\ell, P}$ and $\mathcal{H}^{(2)}_{\ell, P}$ removing the last and first column of $\mathcal{H}_{\ell, P}$ respectively. It can be shown that the generalized eigenvalues of the pair $\left\{ \mathcal{H}^{(1)}_{\ell, P}, \mathcal{H}^{(2)}_{\ell, P} \right\}$ are of the form $e^{j \frac{2 \pi}{M} \omega_{\ell,q}}$, where $\omega_{\ell,q}$ is the spatial frequency associate with the i-th component~\cite{pencil1, pencil2, pencil3, pencil4}.

\bibliographystyle{IEEEtran}
\bibliography{refs}

\begin{IEEEbiography}[{\includegraphics[width=1in,height=1.25in,clip,keepaspectratio]{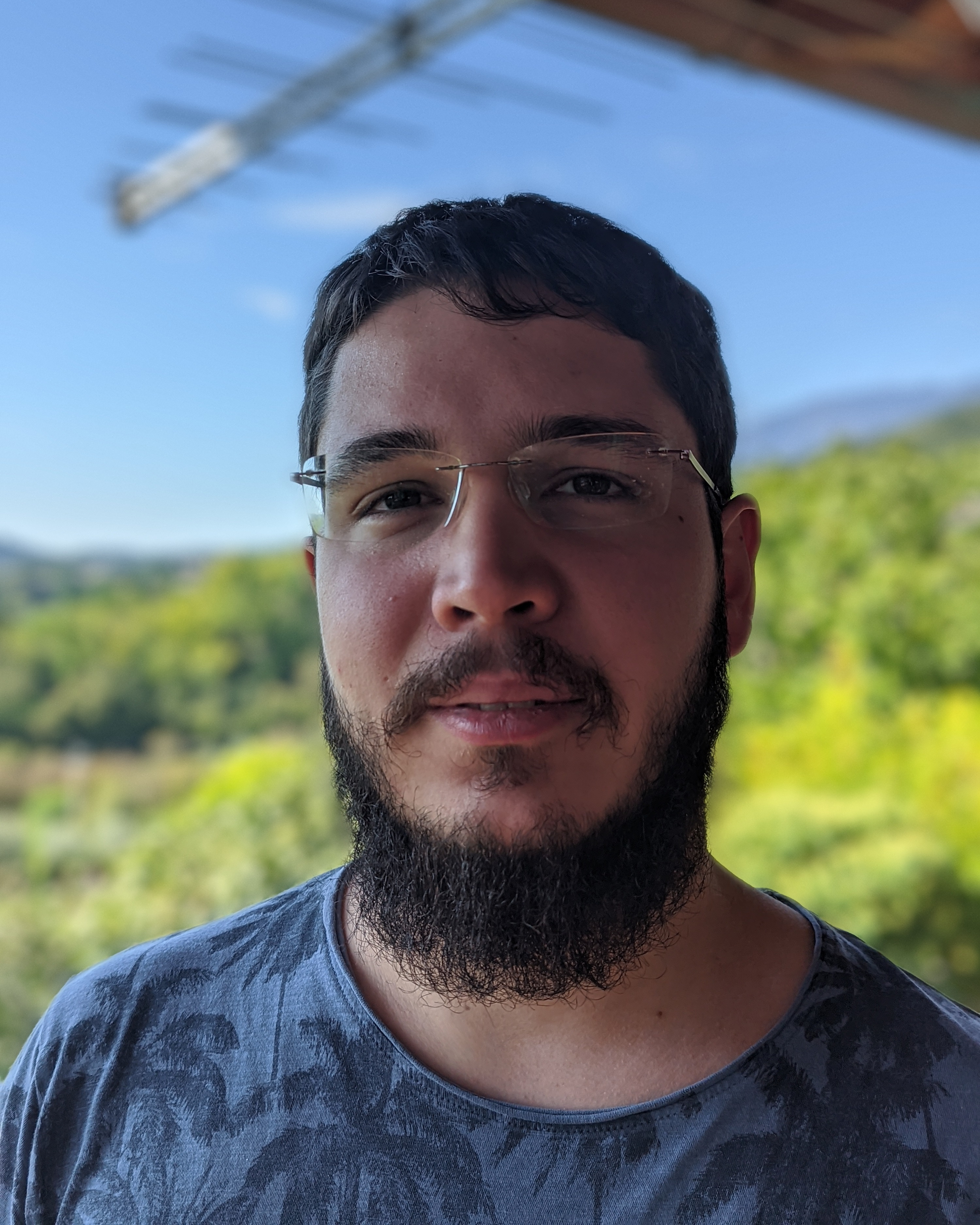}}]{Andrea Bedin} is an Early Stage Researcher within the MINTS project, currently working at Nokia Bell Labs in Espoo, Finland. He is also enrolled in the Ph.D. program in information engineering at the University of Padova. He obtained his master’s degree in ICT for internet and multimedia engineering in 2020 and his bachelor’s degree in information engineering in 2018 from the University of Padova. His research interests are in low latency ultra-reliable communications for industrial applications.
\end{IEEEbiography}

\begin{IEEEbiography}[{\includegraphics[width=1in,height=1.25in,clip,keepaspectratio]{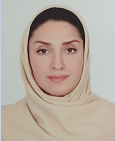}}]{Shaghayegh Shahcheraghi} was a researcher at Technical University of Darmstadt and a member of MINTS project from 2020 to 2022. She received a B.Sc. in Electrical Engineering from Shiraz University, 2008 to 2012, and an M.Sc. in Electrical Engineering-Telecommunication-Field from Shiraz University, 2012 to 2014. She received her second M.Sc. in Telecommunication Engineering-Signal processing from Polytechnic University of Milan, 2017 to 2020. Her research interests include signal processing, join communication and sensing, and positioning.
\end{IEEEbiography}

\begin{IEEEbiography}[{\includegraphics[trim=40 0 40 0,width=1in,height=1.25in,clip,keepaspectratio]{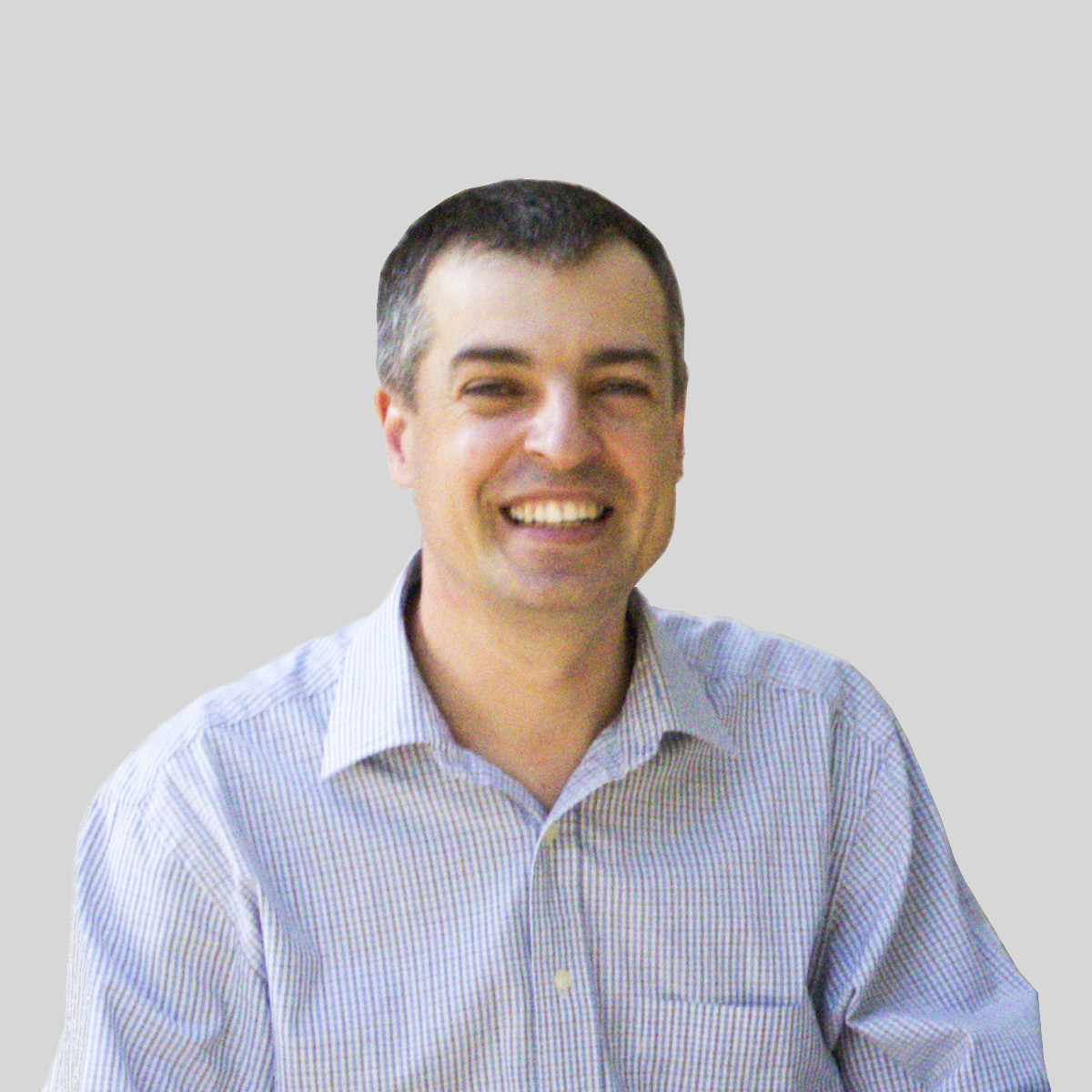}}]{Traian E. Abrudan} is a senior research scientist at Nokia Bell Labs, Espoo, Finland. Previously, he was a postdoctoral researcher with the Department of Computer Science, University of Oxford, UK (2013-2016). During 2010-2013, he was a postdoctoral researcher with the Faculty of Engineering, University of Porto, Portugal, and a member of Instituto de Telecomunica\c{c}\~oes. Between 2001-2010, he was a member of SMARAD (Finnish Centre of Excellence in SMArt RADios and Wireless Research) which has been selected as Center of Excellence in research by The Academy of Finland. Dr. Abrudan received his D.Sc. degree (with honors) from Aalto University, Finland (formerly known as Helsinki University of Technology) in 2008, and the M.Sc. degree from the Technical University of Cluj-Napoca, Romania in 2000. His fundamental research topics include statistical signal processing, numerical optimization and machine learning applied to multi-antenna communication and sensing.
\end{IEEEbiography}

\begin{IEEEbiography}[{\includegraphics[trim=40 0 40 0,width=1in,height=1.25in,clip,keepaspectratio]{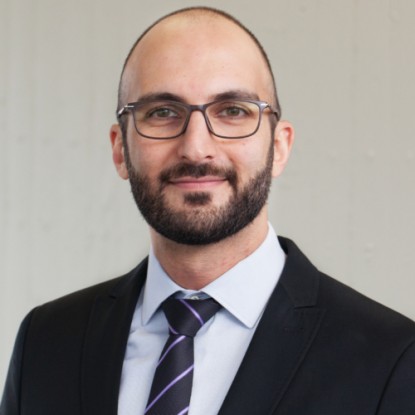}}]{Arash Asadi}  is a research group leader at TU Darmstadt, where he leads the Wireless Communication and Sensing Lab (WISE). His research is focused on wireless communication and sensing and its application in Beyond-5G/6G networks. He is a recipient of several awards, including Athena Young Investigator award from TU Darmstadt and outstanding PhD and master thesis awards from UC3M. Some of his papers on D2D communication have appeared in IEEE COMSOC best reading topics on D2D communication and IEEE COMSOC Tech Focus.
\end{IEEEbiography}

\vfill

\end{document}